\documentclass[fleqn,usenatbib]{mnras}

\usepackage{newtxtext,newtxmath}

\usepackage[utf8]{inputenc}
\usepackage[T1]{fontenc}

\DeclareRobustCommand{\VAN}[3]{#2}
\let\VANthebibliography\thebibliography
\def\thebibliography{\DeclareRobustCommand{\VAN}[3]{##3}\VANthebibliography}

\usepackage{graphicx}	
\usepackage{amsmath}	

\title[Neural Networks for Galaxy Properties]{Not Hydro: Using Neural Networks to estimate galaxy properties on a Dark-Matter-Only simulation}

\author[C. A. Hernández et al.]{
Cristian Hernández Cuevas,$^{1}$\thanks{E-mail: cahernandez1@uc.cl}
Roberto E. González,$^{2}$
Nelson D. Padilla$^{3}$
\\
$^{1}$Instituto de Astrofísica, Pontificia Universidad Católica de Chile, Vicuña Mackenna 4860, Macul, 8970117, Santiago, Chile\\
$^{2}$Centro i+d EY MetricArts, Presidente Riesco 5435, 4° Floor, Las Condes, 7550000, Santiago, Chile\\
$^{3}$Instituto de Astronomía Teórica y Experimental (IATE), CONICET-UNC, Laprida 854, X5000BGR, Córdoba, Argentina
}

\date{Accepted XXX. Received YYY; in original form ZZZ}

\pubyear{2022}
\begin{document}
\label{firstpage}
\pagerange{\pageref{firstpage}--\pageref{lastpage}}
\maketitle

\begin{abstract}
Using data from TNG300-2, we train a neural network (NN) to recreate the stellar mass ($M^*$) and star formation rate (SFR) of central galaxies in a dark-matter-only simulation. We consider 12 input properties from the halo and sub-halo hosting the galaxy and the near environment. $M^*$ predictions are robust, but the machine does not fully reproduce its scatter. The same happens for SFR, but the predictions are not as good as for $M^*$. We chained neural networks, improving the predictions on SFR to some extent. For SFR, we time-averaged this value between $z=0$ and $z=0.1$, which improved results for $z=0$. Predictions of both variables have trouble reproducing values at lower and higher ends. We also study the impact of each input variable in the performance of the predictions using a leave-one-covariate-out approach, which led to insights about the physical and statistical relation between input variables. In terms of metrics, our machine outperforms similar studies, but the main discoveries in this work are not linked with the quality of the predictions themselves, but to how the predictions relate to the input variables. We find that previously studied relations between physical variables are meaningful to the machine. We also find that some merger tree properties strongly impact the performance of the machine. 
We conclude that ML models are useful tools to understand the significance of physical different properties and their impact on target characteristics, as well as strong candidates for potential simulation methods.
\end{abstract}




\section{Introduction}

Cosmological simulations are a powerful tool to understand the nature and evolution of galaxies, large-scale structures and baryonic processes occurring within them. The $\Lambda \text{CDM}$ universe can be described with only six observationally-tuned parameters 
, and the evolution of the cold dark matter in the universe is modelled by following gravitational interactions between dark matter. The complexity of the models only increases when one takes in consideration the physics of baryonic matter. Simulations follow these models with the goal of replicating the evolution of the large-scale universe and the emergence of different bodies and structures. 

The first simulations created were dark-matter-only (hereafter DM-Only), only having gravitational interactions ~\citep{aarseth_1971} between particles. Simulations that model the evolution this way are referred to as N-body simulations. With time, simulations evolved into more sophisticated algorithms, such as semi-analytical models, that coupled to DM-Only simulations were able to follow barionic properties of galaxies (hereafter SAM) \citep{1993MNRAS.264..201K}, and hydrodynamical simulations (hereafter hydro) simulations ~\citep{1992ApJ...399L.109K}. 
The main difference between SAM and hydro simulations is that SAMs take the approach of using approximate, analytic techniques to treat the various physical processes associated with galaxy formation, which makes them computationally cheaper \citep{cole2002}. In comparison, hydro simulations solve the physics of galaxy formation by computing directly the fundamental equations of gravitation, hydrodynamics, cooling and star formation, and in some cases even radiative 
transfer between a large number of particles. Both SAM and hydro simulations can be contrasted with observations, thus testing the physical processes involved from different perspectives.

The purpose of simulations is to understand the physical processes involved in the formation and evolution of galaxies and the universe. Since some simulations are more complex than others, there is a large variety of scales and objects to study: from star formation and evolution to galaxies and the large-scale structure of the universe. Simulations have different applications. For example, pre-analyzing large future observational surveys of galaxies ~\citep{sanchez2021sniacosmology,2021sim,2019sim}. The majority of cosmological surveys are focused on observing galaxies, therefore for a simulation it is vital to consistently reproduce different features of galaxies and baryonic matter. However, the modeling of said galaxies is non-trivial due to the complex physical processes in their formation and evolution. For even a small fraction of the Universe -in the context of hydrodynamical simulations- evolving tens of billions of  particles interacting under coupled effects of gravity, magneto-hydrodynamics and radiative processes over cosmic time is incredibly computationally costly. This cost increases with the volume of the desired simulation. For example, TNG300-1 (one of the simulations of the suite IllustrisTNG ~\citep{nelson2021illustristng,Pillepich_2017,illu1,illu2,illu3,illu4}), which has a simulation box of 205Mpc$/h$ per side, required almost 35 million CPU hours to complete \citep{yip2019dark,Pillepich_2017}. This poses a challenge, since new surveys are incrementally larger and simulations must keep the pace with such large volumes of data ~\citep{2013skysurveys}.

In contrast to magneto-hydrodynamical simulations, DM-Only N-body simulations are computationally much cheaper as gravity is the only interacting force. For example, the DM-Only simulation Millennium, which traces the evolution of dark matter in a cube of roughly 500Mpc$/h$ on a side, took only 350 thousand CPU hours \cite{2005mi}. In contrast with TNG300-1, Millenium took 100 times less computing time for a volume 14 times larger. On top of this, we must consider that finding haloes and caracterizing them is already a time-consuming task \citep{Knebe_2011}. Therefore, it would be extremely interesting to find a time-efficient mapping from dark matter characteristics in N-body simulations to the baryonic properties in full hydrodynamical simulations.

With this in mind, the goal would be finding this mapping to learn from a different perspective the variables that govern the galaxy formation process.
While we know statistical relations between dark matter properties and baryonic features, we propose a different approach. Since simulations cover very large volumes, there is a flood of information to process and find the mapping we seek. Since  DM-Only simulation can provide merger trees as their output, we can somewhat reduce the complexity of the whole simulation by tracking particles in said trees and inferring the dark matter properties of haloes and subhaloes that emerge. In a typical simulation, we can find hundreds of thousands of subhaloes ~\citep{2016eagle,2021countsub, https://doi.org/10.48550/arxiv.1607.03224, Dolag_2009, G_mez_2021},  with a set of features like mass, half mass radius, spin, velocity dispersion, etc. Therefore, the "domain" of this mapping contains a very large amount of data.

There are quite a few theoretical models which describe different relationships between the dark matter environment of baryonic matter and its properties, using approaches with a phenomenological approach \citep{2013sfrhalo}, simulation-based \citep{2018sfrobs} or via machine learning ~\citep{Jo_2019,Agarwal_2018} methods. But with the amount and variety of data we are trying to model we would need a quite complex model to describe the relationships between all features of interest. In the spirit of understanding large volumes of data, machine learning (hereafter ML) is a strong alternative to traditional methods. Even more so, supervised ML specializes in constructing mappings between a set of measurements (input) and a target variable (output). There is one condition though. In order for the algorithm to "learn", it needs a set of provided examples. Furthermore, a larger  set of inputs and outputs may improve the performance of the mapping fit by the algorithm ~\citep{sun2017revisiting,2015data}. Once obtained, the mapping function can be used to predict the output of previously-unseen inputs. The main difference between traditional modeling and supervised ML is that the mapping is predefined in traditional methods, while the supervised algorithm constructs the mapping according to the training data.

A second advantage for ML algorithms is the computing time as most of the computing resources are needed in the training phase of the algorithm. After that, the cost of inference of an output by a trained machine is low. There are quite a few fields where ML algorithms outperform classical methods used for the same means. In astrophysics there has been a rapid increase of studies on the applications of ML methods to process different types of data. Examples range from detection, classification and analysis of structures in astronomical images ~\citep{gonzalez2018galaxy,2017jacobsDL} and spectrographic data ~\citep{2016spectra}. The application of ML in this area is not only oriented to observational data. Studies have been successful in using ML methods to analyze, predict and even replicate data from astrophysical simulations \citep{2016mlsim, Agarwal_2018}. The nature itself of astronomical data (both in form and volume) make astrophysics and ML kindred fields of study.

For simulations to be useful for the prediction and analysis of cosmological surveys, one must take in account the nature of the surveys. Some surveys detect galaxies using photometric observations (e.g. LSST \citep{2019ApJ...873..111I}). To replicate these observations one must simulate properties such as the stellar mass of galaxies properly. Other surveys detect galaxies by observing the line emission spectra (e.g. DESI \citep{desicollaboration2016desi}, EUCLID \citep{laureijs2011euclid}). In this case, line emission responds to the ionizing flux. Therefore, properties like star formation rate must be properly simulated since it is massive, short lived stars the ones that produce most of this flux
\citep{orsi2014}.

In this work we train a neural network (hereafter NN) to find a mapping from dark matter data to baryonic target variables. We train the algorithm using the publicly available catalog of IllustrisTNG, more specifically the TNG300-2 simulation. Our goal is to fit a mapping from a selection of variables of the DM-Only simulation data to two target baryonic properties found in the complete magneto-hydrodynamical simulation: stellar mass and star formation rate. We will use dark matter properties from the halo and subhalo containing a galaxy to infer its stellar mass and SFR. For SFR in particular, we time-averaged this value between consecutive snapshots $z=0$ and $z=0.1$, with the idea of reducing stochasticity of this variable at $z=0$ and getting a better idea of SFR as it changes in time and not as a strictly instantaneous value. This averaging will improve our final results. This study will focus on central galaxies of haloes, ignoring galaxies contained in satellite subhaloes. In line with the nature of the methods employed, this work focuses in the influence of the data from the dark matter environment on the target galaxy properties, and not necessarily on the comprehension of the physical processes behind said influence. We study the importance of input variables on the performance of the model using a leave-one-covariate-out approach, which led to insights about the relation between input variables; in particular those that are heavily correlated. To evaluate the performance of the algorithms, we will use mean squared errors (MSE), Pearson Correlation Coefficient (PCC) and Coefficient of determination ($R^2$) metrics. On top of that, we will study the distribution of the predicted and real values of the target variables to be regressed and compare to similar studies in literature.

Details about the simulation and the variables involved in the training are discussed in Section 2. In Section 3 we will discuss the theoretical background between the relationship of stellar mass and SFR of galaxies with the dark matter environment, and talk about work related to ours. In Section \ref{chap:ML} we will discuss the nature of the ML algorithms used and its modeling process. In Section \ref{chap:results} we will present the results of our work. We will discuss and analyze these results in Section \ref{chap:discussion}. Finally, in Section \ref{chap:discussion} we present the conclusions of our work.

\section{Cosmological Simulation and Data}\label{chap:sim_data}

In this work we use the IllustrisTNG suite of large volume, cosmological, gravo-magneto-hydrodynamical simulations that model the physical processes most relevant to the formation and evolution of galaxies in cosmological volumes \citep{2017}. These simulations were run with the moving-mesh code AREPO ~\citep{Springel_2010} to solve the magneto-hydrodynamics and self-gravity coupled equations ~\citep{nelson2021illustristng}. This moving-mesh code employs a tree-particle-mesh algorithm to solve Poisson’s equation for gravity and a second-order accurate finite-volume Godunov scheme on a moving, unstructured Voronoi mesh for the equations of ideal magneto-hydrodynamics \citep{Pillepich_2017}. The TNG project is made up of three flagship runs, each with different volume and particle resolution: TNG50, TNG100, and TNG300. For this work, we will focus on TNG300, the largest simulation. It has a volume of roughly 300 $cMpc^3$. It is presented in three versions. From highest resolution to lowest, they are: TNG300-1, TNG300-2 and TNG300-3. Our work uses data from TNG300-2. This simulation took 1.3 million CPU hours on 6000 cores \citep{Pillepich_2017}. At $z=0$, the simulation volume holds over two million subhaloes, identified with the SubFind \citep{Springel_2001} and friends-of-friends (FOF \cite{1985ApJ...292..371D}) halo finders. This simulation includes all relevant galaxy-scale physics to follow the evolution of dark matter, stars, gas, and super massive black holes.
 
Each simulation of the TNG suite, and specfically TNG300-2, have a DM-Only counterpart. They are run with the same initial conditions of their magneto-hydrodynamic counterpart, but only with dark matter particles \citep{nelson2021illustristng}. Since they share initial conditions, the haloes and subhaloes that emerge are quite similar. On top of that, there is a cross-match subhalos between baryonic and dark matter runs.
Said cross-match is a data product from \cite{2015submatch} for SubLink \citep{sublink} found subhalo matching. There also exists a match for LHaloTree \citep{LHalo2015}. This latter match is not used in this work.

\subsection{Contrast between TNG300-2 and observational data} \label{sec:contrast}

The public data release paper of IllustrisTNG \citep{nelson2021illustristng} reports consistent results in several aspects when compared with observations. For example, \cite{dmfraction2018} finds that the dark matter fraction (DMF) at $z=0$ falls among estimates for disk-like galaxies from the SWELLS and DiskMass samples from the SDSS survey. They also find that, for Milky-Way-Like galaxies, the total circular velocity curves beyond a few kpc from the galaxy centre behave accordingly with observational constraints. Elliptical galaxies show DMF in agreement with the measurements made from the SLUGGS survey by \cite{2013MNRAS.428.2407W}, but higher than the measurements from \cite{dmf10.1093/mnras/stx678}. Another relation found by \cite{Pillepich_2017} is that IllustrisTNG reproduces the general features of the stellar mass - halo mass relation (SMHM) semi-empirical constraints (as seen in \cite{simval2016} and \cite{2017}). This work also finds that, while the total simulated amount of stellar mass in clusters is in agreement with available observational values, the mass in central galaxies appears up to a factor of 0.5 dex larger than the observational constraints in \cite{stcons2018}.

With respect to the SFR, it is treated following \cite{sfrhow}, where gas cells are stochastically converted into star particles using a density threshold criteria. Gas cells with $n_H>0.1cm^{-3}$ are considered to be \textit{star forming}. The nature of this SFR is instantaneous, and the SFR of a galaxy is measured by summing the instantaneous SFR of all its star forming gas cells  \citep{sfrtng2019}. Due to resolution issues, any gas cell in TNG300 with $log(SFR)<-3$ is considered as unresolved and assigned a SFR value of 0. When contrasted with observational data, we can see at $z=0$ that the threshold to select star-forming v/s quenched galaxies in the UVJ diagram in \cite{quenchedsfr} can be reasonably well applied to TNG galaxies to separate in a consistent fashion red, quenched galaxies from blue, star-forming ones. With this said, the TNG galaxies populate the UVJ diagram in a broadly successful way, but not identical to observations. On top of that, TNG succesfuly recreates the quenching at high stellar masses, since massive galaxies tend to be older and non star forming \cite{sfrtng2019}. In the aforementioned work, we can also see that TNG galaxies populate the $SFR-M_{star}$ plane in a qualitatively consistent fashion with observations. With this said, due to the different nature of SFR observational indicators, the authors limit their comparison between TNG and observations by only focusing on the slope and mass trends of the star-forming main sequence. While the agreement with observations falls short at high redshifts, at $z=0$ the main sequence of TNG galaxies lies inside the range of observational constraints bracketed by the measurements of \cite{oliver} and \cite{zahid}. Finally, as previously mentioned, the SFR is calculated instantaneously which is not a factually observable measurement. Therefore, to study SFR in a way that makes sense when compared to observations, \cite{sfrtng2019} propose averaging SFR over some timescale. They find that longer averaging time-scales lead to smaller levels of scatter. At low redshifts ($z<2$), and by accounting for measurement uncertainties in stellar mass and SFR, the main sequence scatter is overall consistent with observational findings \citep{scatterSFR}.

In general, TNG simulations consistently match observational relations and constraints. Even though we see that some features do not perfectly match the empirical data, we can see a plethora of factors that make TNG a reliable model which closely resembles observable attributes of haloes and galaxies; thus making it a trustworthy source of simulated astrophysical data.

\subsection{Physical Models and Numerical Methods}

The IllustrisTNG simulations assume a cosmology consistent with the Planck Collaboration ~\citep{2016} results: $\Omega_{\Lambda,0} = 0.6911$, $\Omega_{m,0} = 0.3089$, $\Omega_{b,0}  = 0.0486$, $\sigma_8 = 0.8159$, $n_s = 0.9667$ and $h = 0.6774$. It assumes Newtonian self-gravity, solved in an expanding Universe i.e. in a cosmological background ~\citep{nelson2021illustristng}. The simulation starts at $z=127$ and runs until $z=0$. At $z=127$, the initial conditions of TNG300-2 consist of 1250\textsuperscript{3} DM particles with $m_{DM} = 470 \times 10^6 M_\odot$ and 1250\textsuperscript{3} gas cells with $m_{gas} = 88 \times 10^6 M_\odot$.

Baryonic TNG runs include additional physical components, including feedback, seeding and growth of supermassive black holes and pressurization of the interstellar medium. Other relevant components for this work are  stochastic star formation in dense interstellar medium gas above a threshold density criterion, and evolution of stellar populations, with associated chemical enrichment and mass loss \citep{nelson2021illustristng}. The details on the behaviour and validation of the physical models are presented in  \cite{physval2017} and \cite{simval2016}.

\subsection{Identifying cosmological structures}

The data product of each simulation is divided in 100 snapshots, each at a different redshifts. At every snapshot, two types of group catalogs are provided: haloes, identified and catalogued by the friends-of-friends (FoF) algorithm, and subhaloes identified with the SUBFIND algorithm \citep{Springel_2001}. 

FoF places any two particles with a separation less than some linking length \textit{b} into the same group. In this way, particle groups (or haloes) are formed, corresponding to regions approximately enclosed within isodensity surfaces with density inversely correlated with the volume of the sphere of radius \textit{b}. For an appropriate choice of b, groups are selected that are close to the virial overdensity predicted by the spherical collapse model \citep{spherical}. This simulation uses a linking length of $b = 0.2$ \citep{nelson2021illustristng}. The FOF algorithm is not capable of detecting substructures inside larger virialized objects with a linking length of this value \citep{Springel_2001}. 

To identify subhaloes, the SUBFIND algorithm is run over the FOF groups data. The use of FOF-groups as input data provides a mean to organize the groups in a simple two stage hierarchy consisting of ‘background group’ and ‘substructure’. This algorithm first identifies overdensities within a given, FOF group. It begins agglomerating neighbour particles to the most dense particle in the vicinity, and setting the subhalo boundaries with criteria based on the density gradient. To switch from a criteria based only on the spatial distribution of particles to a more physical definition, a requirement of self-boundedness is set. This is done by removing any particle with positive energy, which are considered unbound. The unbinding is performed in physical coordinates, where velocities (and therefore energy) are computed by using the most bound particle (the one with less potential energy) as the center. \citep{Springel_2001}.

\subsection{Halo catalogues} \label{sec:product}

Among the group catalogs generated by the FoF algorithm, and the subhalo catalog generated by the SUBFIND algorithm, the release of IllustrisTNG also makes available 100 snapshots which contain data for every particle and cell in the whole volume. Each snapshot captures the state of galaxies, haloes, particles and cells at different refshifts. 
In this work we will use data from two different full snapshots at redshifts  $z=0$ and $z=0.1$.

Another data product of this simulation are the merger trees. Merger trees are a data structure which follow the growth and mergers of dark-matter haloes over cosmic history. These give important insights into the growth of cosmic structure, allowing to trace the history of the dark matter interactions involved during the formation of a halo or subhalo \citep{merger1}. IllustrisTNG has available merger trees created using SubLink \citep{sublink} and LHaloTree \citep{lhalotree}. Merger trees will allow us to trace back on time major mergers (mergers between haloes with a mass ratio of 1/3) and the past mass of subhaloes at different redshifts.

\subsection{Input and Output Data}

For this work, we focus on present day ($z=0$) central subhaloes. In order to train the machine we will use properties from the subhalo, the halo hosting this subhalo, neighbour halos, and merger history of subhaloes. This data is obtained from the DM-Only version of TNG300-2. At the same time, we take three baryonic properties of subhaloes from the baryonic TNG300-2 run, which are linked to the DM-Only data by the previously mentioned cross-match catalog. The input data we use in our algorithms, from the TNG300-2 DM-Only simulation are described in Illustris' web page\footnote{\href{https://www.tng-project.org/data/docs/specifications/}{IllustrisTNG Data Specifications}.} are:

\subsubsection{Subhalo properties}

We use subhalo properties as these have been shown to be highly correlated to galaxy properties such as the stellar mass \cite{Rodr_guez_Puebla_2016}.

\begin{enumerate}
    \item $S_\text{sub}$ ($\frac{kpc}{h}\frac{km}{s} $): Magnitude of the spin of the subhalo, computed for each as the mass weighted sum of the relative coordinate times relative velocity of all member particles/cells.
    \item  $\sigma_\text{sub}$ ($\frac{km}{s}$): One-dimensional velocity dispersion of all the member particles/cells in the subhalo (the 3D dispersion divided by $\sqrt{3}$).
    \item $v_{max}$ ($\frac{km}{s}$): Maximum value of the spherically-averaged rotation curve, i.e. maximum circular velocity of the subhalo.
\end{enumerate}
\subsubsection{Host halo properties}

We also include host halo properties since these can be of importance to central galaxies.  If we were to include satellites, the halo properties would also come into play as environmental properties.

\begin{enumerate}
    \item $m_{halo}$ ($log(M_{\odot}/h)$): Logarithm of the sum of the individual masses of every DM particle in the halo.
    \item $r_{crit,200}$ ($ckpc/h$): Comoving Radius of a sphere centered at the most bound particle in the halo with a mean density of 200 times the critical density of the Universe, at $z=0$.
    \item $r_{crit,500}$ ($ckpc/h$): Comoving Radius of a sphere centered at the most bound particle in the halo with a mean density of 500 times the critical density of the Universe, at $z=0$.
    \item $m_{crit,200}$ ($10^{10}M_{\odot}/h$): Total Mass of this halo enclosed in a sphere whose mean density is 200 times the critical density of the Universe, at $z=0$.$m_{stellar}$
\end{enumerate}
\subsubsection{Environment properties}

We include environmental properties as these can be related to historical events in the evolution of a galaxy. See Section \ref{sec:theory}.

\begin{enumerate}
    \item $\rho_n$ ($(ckpc/h)^{-3}$): Numerical density of neighbour haloes, computed as $5/V_5$, where $V_5$ is the volume of the sphere with a radius equal to the distance to the fifth closest halo.
    \item $\rho_{mass}$ ($(10^{10}M_{\odot})(ckpc/h)^{-3}$): Mass density of halo neighbourhood, computed as $m_5/V_5$, where $V_5$ is the volume of the sphere with a radius equal to the distance to the fifth closest halo and $m_5$ is the sum of the masses of the five closest haloes.
\end{enumerate}

\subsubsection{Historical properties}

Related to the environmental properties, we can also include historical properties directly in the analysis to then compare their relative influence.

\begin{enumerate}
    \item $z_{1/2}$: Redshift at which this subhalo had half of its actual mass.
    \item $\dot{m}_{subhalo}$ ($10^{10}M_{\odot}/Gyr$): Free dark matter particles accreted by the subhalo from $z=0.1$ to $z=0$
    \item $z_{last}$: Redshift at which this subhalo had its last major merger. This is, when it merged with another subhalo such as their mutual mass ratio is at least 1/3.
\end{enumerate}

\subsubsection{Output features}

On the other hand, we choose two output galaxy properties that come from the TNG300-2 magneto-hydrodynamical simulation.  These properties are fundamental and key in the selection process of surveys.  In addition we also include a variant of one of the two, that will come in handy for our analysis in later sections:

\begin{itemize}
    \item $m_{*}$ ($log(M_{\odot}/h)$): stellar mass obtained as the sum of the masses of all star and wind particles within twice the stellar half mass radius.
    \item SFR ($M_{\odot}/yr$): star formation rate obtained as the sum of the individual star formation rates of all gas cells within twice the stellar half mass radius. Instantaneous measure.
    \item meanSFR ($M_{\odot}/yr$): Time-scale averaged SFR between $z=0$ and $z=0.01$. As mentioned in \S\ref{chap:background}, this measure is a better representation of the observationally measured SFR and presents less scatter than the instantaneous SFR.
\end{itemize}

\section{Theoretical Background}\label{chap:background}

\subsection{Galaxy properties and the relation to their host haloes and environment} \label{sec:theory}

As the universe evolves, baryonic and dark matter interactions shape the structures we are able to observe today. Since the dark matter only interacts through gravity, and it dominates the gravitational potential in the universe, it is safe to assume that its properties and the ones of the baryonic matter have an impact on each other. Several studies have found correlations between the properties of haloes and the galaxies that inhabit them. For example, studies have found strong correlations between measured halo mass and the observed stellar mass of the galaxies they host \citep{halostar1,halostar2,2013sfrhalo,smhm2020}. 

In the literature there are records of a strong relation between SFR and halo mass, although this relation is mediated by (i.e. more related to) the stellar mass of the galaxy \citep{2013sfrhalo,sfrtheo1,sfrtheo2,sfrtheo3,sfrtheo4}. 
This relation is known to evolve with redshift.  Numerical studies and observational estimates show that on top of that, the past events of haloes also have an impact on the SFR, where mergers show an elevated SFR in comparison to similar non-merging galaxies \citep{sfrmerg1,sfrmerg2,sfrmerg3}.  Since they both respond to the gravitational potential, the features of the dark matter of a halo should be able to characterize, one way or another, its hosted galaxy. But finding relations between a high number of features requires several observational data and complex models of coupled equations. But, if said relations exist, one avenue should be  to describe them with a physically motivated model and this has been done extensively \citep{cole2002,sage2016,mdplsag}. 

Instead of constructing a model this way, in this work we use machine learning methods to implicitly infer the underlying relations between halo and galaxy properties. This approach will not provide us an analytical model to understand said relations, but it will allow to study the impact of the different halo features in the galaxy properties. There are machine learning methods that ideally require large amounts of pre-analyzed data. This poses a challenge if we intend to learn from observations, because on top of needing to process these, training on data from different datasets (e.g. telescopes) can result in different predictions \citep{JMLR:v9:crammer08a}. For this reason we apply these methods to data not from observations, but from the cosmological simulation IllustrisTNG.

\subsection{Previous ML related work} \label{sec:related}

Even though there are several works studying the performance of ML to infer galaxy and baryonic properties from DM only simulations \citep{Kamdar_2015, Villaescusa_Navarro_2021, de_Santi_2022}, we will concentrate our comparisons to two studies, previously cited in this work, and highlight differences and similarities with them. In first place, we have \cite{Agarwal_2018} work. They also developed a ML framework to infer baryonic properties such as Metallicity (Z), neutral ($H_1$) and molecular ($H_2$) hydrogen, and our variables of study: SFR and stellar mass. They use the hydrodynamical simulation MUFASA \citep{mufasa2016}, which is smaller and has less resolution than TNG300-2. They explore a few ML algorithms, among them Multi-layer perceptrons (MLP hereafter). While the scatter of the relationship between output variables and halo mass was underpredicted, they recovered the mean trends of output quantities with halo mass highly accurately. In their work they didn't get the best results using MLP but using Random Forest. Also, they study the impact of additionally inputting key baryonic properties (like stellar mass or SFR) when predicting $H_1$ and $H_2$, as would be available e.g. from an equilibrium model. They found that in doing so, their results improved. This result inspired us to use ML inferred baryonic properties to improve the performance of our models, as will be detailed in Section \ref{chap:ML}. We will also compare the metrics of the regressions obtained in this work with ours. The results from their regressions, which we compare later with our results, are shown in Table \ref{table:agarwal}.

\begin{table}[b]
        \centering
    \begin{tabular}{|c|c|c|}
    \hline
          & stellar mass & SFR \\
         \hline
        $R^2$ & 0.909 & 0.555 \\
        PCC & 0.953 & 0.745 \\
        \hline
    \end{tabular}
    \caption{Metrics of best regressions for stellar mass and SFR in \citep{Agarwal_2018} at $z=0$}
    \label{table:agarwal}
\end{table}

The second related study is \citep{Jo_2019}. In this work they employ ML methods to estimate baryonic properties of a galaxy inside haloes from a DM-only simulation. They work with TNG100, a smaller simulation with higher resolution than TNG300-2. They train a machine to predict features like stellar mass and star formation rate in a galaxy based on the DM content of the halo that hosts it. The ML algorithm used by them is Extremely Randomized Trees \citep{ERT}, a variation of the Random Forest algorithm. For a baseline training, they use only 3 properties of the halo: DM mass, velocity dispersion and maximum circular velocity of the halo. Then, they use different approaches to improve their results.  One approach is augmenting the baseline dataset. They add the halo spin, historical properties like number of mergers and last major merger mass ratio, and environmental properties like local density of haloes and number of local halos. Another approach involves a two-stage learning procedure, where they use a first machine to predict a baryonic property. This property is then used as an input to train a second machine. One last approach is to use an error function with logarithmic scaling. While different combinations of approaches sometime interfere with each other, they find that using adequate combinations (which are different for each baryonic property) the results improve. Once they find the best machine for each property, they generate a galaxy catalog with the studied baryonic properties for another DM-Only simulation: MultiDark-Planck \citep{multidarkplanck1}\citep{multidark2} \citep{multidark3}. Finally, they compare the machine's performance against semi-analytic model (SAM) data, the MDPL2-Sag catalogue \citep{mdplsag}. They compare the probability distribution function (PDF) of each baryonic property between TNG100, SAM (Sag) and their machine. Overall, they find that while the machine replicates better the PDF of TNG100 (which it was trained to do), there are some clear mismatches in some higher or lower ends of the distributions of properties, reported to be due to small number statistics. In summary, they found that adding environment and historical properties and employing a two-stage learning method improves their results, and that a catalog generated by their method is largely compatible with a SAM catalog.

\section{Machine Learning Methods and Modeling}\label{chap:ML}

In this Section we present our machine learning setup.

\subsection{Supervised Learning}

Supervised learning algorithms are trained to find complex relations from previous data. They must be provided with a set of input-output pairs. For example, in this work the input data corresponds to the DM-Only features while the output data are the two baryonic properties: stellar mass and SFR.

The machine tries to learn the best mapping from inputs to outputs, so that when a new input (from which the machine didn't learn) is given to the machine it outputs a prediction based on the data used to learn said mapping. To learn, the machine begins answering randomly and it iteratively improves its answers by adjusting its weights under an optimization scheme to reduce a given loss function, i.e. gradient descent algorithms. The data used by the algorithm to learn the optimal mapping is called training set. 
To evaluate how good is the mapping learned by the machine, we must take a set of data which the machine has not "looked at" before (i.e. it is not in the training set). By taking data for which we know the output, we can compare the machine's prediction with the real value and use mathematical methods (metrics) to evaluate the performance of the trained machine. The set of data used to evaluate the performance of the machine is called validation set.

Apart from the internal parameters previously mentioned, supervised algorithms have a set of parameters which must be previously given by the user (called hyperparameters). These hyperparameters are tuned to achieve the best metrics when evaluating in the validation set. Once we find an optimal set of hyperparameters, the final performance of the machine is evaluated on a third set of data that is different from both the training and validation set. This set of data is called testing set. To begin training an algorithm, one usually divides all the data available in three mutually exclusive sets: training, testing and validation.

\subsection{ML Setup} \label{sub:setup}

For this work we build custom MLP using the \textit{keras} and \textit{tensorflow} packages for machine learning in a \textit{python} script. We train two MLP, one for each output. In early stages of the development of this research, both outputs were predicted using only one model, but this proved to be inefficient metric-wise and it reduced the volume of data due to the restrictions mentioned in section \ref{sec:constraints}. While we will explore the behaviour of the machine in various methods for predicting SFR, we will attune the machine for meanSFR since it is a more significant value from a physical perspective, as mentioned in section \ref{sec:contrast}. For short, we will address this machine as the SFR machine. After thoroughly exploring the hyperparameter space and architectures, we chose the number of hidden layers, neurons per layer, learning rate and batch size for training. Each variable to be regressed has a different machine, i.e. with different hyperparameters. The hidden layer neurons use a ReLU activation function \citep{pmlr-v15-glorot11a}, and the output layer uses a softplus activation function \citep{softplus}. We use an ADAM optimizer \citep{kingma2017adam} in both cases. We train using mean squared error as the loss function, since it gives a harsher penalization on large errors.

The MLP for the stellar mass prediction has 3 hidden layers of 30 neurons each. The learning rate for the Adam optimizer has a value of $\text{5}\times\text{10}^\text{-4}$. The machine was trained for 20 epochs using a batch size of 128.

On the other hand, the SFR machine has 4 hidden layers with 40 neurons each. The learning rate for the Adam optimizer has a value of $\text{5}\times\text{10}^\text{-4}$. The machine was trained for 22 epochs using a batch size of 64.

\subsection{Hyperparameters, width and depth exploration} \label{sub:hyper}

\begin{figure*}

    \centering
    \includegraphics[width = 0.95\textwidth]{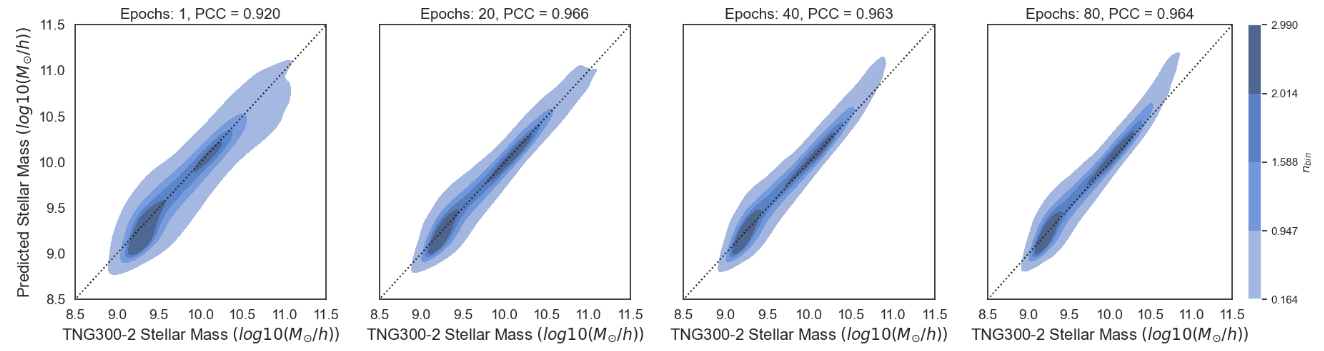}
    
    \caption{Kernel density estimate (KDE) plot of four stellar mass predictions on the testing set using the same hyperparameters at different epochs to illustrate the evolution of the performance of the predictions. On the first panel from left to right it can be seen that the fit is already reasonably good after the first epoch of training. In the second panel the fit has evolved and is closer to the perfect fit shown as the dotted line. The third and fourth plots show that the predictions become less accurate  due to overfitting to the training set. These panels illustrate how we explore for the optimal number of epochs.}
    \label{fig:epochs}
\end{figure*}

To determine the optimal learning rate, batch size, number of hidden layers and number of neurons per layer; we intensively explored combinations of hyperparameters. For each combination, we run five trainings with different random seeds at a fixed number of epochs. Once a training finishes, we evaluate metrics on the validation set and record those values. Metrics are chosen by studying boxplots of the performance of each hyperparameter. Then we train using the chosen combination for different epochs, evaluating how the distribution of predicted versus real values behaves. After that, we repeat the hyperparameter exploration in a smaller range of values around the previously found best values, using the lowest number of epochs where a good fit was observed. We repeat this back-and-forth exploration iteratively until results converge. 
In figure \ref{fig:epochs} we see the evolution of the distribution of real versus predicted values for different epochs for the stellar mass prediction on the testing set. There is a clear improvement with smaller dispersion and better predictions at 20 epochs, and we can see in the figure that for more epochs the correlation barely changes, and the distribution of predicted values underestimates stellar mass for higher TNG values. In particular, when exploring the optimal number of epochs, we tested with values as high as 200 epochs, but the model began converging towards stable metrics at near 20 epochs in both cases.

\subsection{Data selection and preparation} \label{sec:constraints}

Given the resolution of the simulation and the nature of the predicted value we apply some cuts to the simulation data to construct our datasets. In the first place, we only consider galaxies contained in haloes with $m_{halo}>10^{11}m_{\odot}$. As mentioned in section \S\ref{chap:sim_data}, the resolution of dark matter particles in TNG300-2 is $m_{DM}=470*10^6m_{\odot}$. This means we are considering haloes with at least ~200 DM particles. Setting a threshold on masses is a resolution-based criteria for preprocessing also used in \cite{Jo_2019} and \cite{Agarwal_2018}. Then, we make two additional cuts; one for predicting stellar mass and the second for SFR. For stellar mass, we will make predictions on galaxies with $M^*>10^9m_{\odot}$, which correspond to about 100 stellar particles per galaxy in TNG300 and is set as the criterion for the minimum stellar mass values for haloes in \cite{Pillepich_2017}. As for SFR, the only cut we apply is that it must be greater than 0, given that TNG has its own threshold to consider a group of gas cells to be star forming ($\text{SFR}<10^{-3}m_{\odot}/yr$ are treated as quenched, i.e. $\text{SFR}=0$) \citep{sfrtng2019}. For meanSFR, we use the same cuts, allowing galaxies with $SFR_{z=0.1} = 0$. As mentioned in \S\ref{sec:contrast}, the portion of quenched galaxies in TNG is consistent with observational parameters, so these cuts should not pose a problem if one intends to use this method to populate synthetic catalogs using data from dark matter only simulations.

On top of the aforementioned cuts, we also performed a min-max normalization on the input data using the \textit{MinMaxScaler} function for the \textit{sklearn} package in \textit{python}. This is a standard procedure when preprocessing input and output data, since it improves performance and reduces propagation errors in ML \citep{589532}.

\subsection{Regression performance criteria and metrics}
\label{subsec:metrics}

In this subsection we present the metrics adopted throughout.

\subsubsection{Mean Squared Error}

The mean squared error (MSE) is the first metric we use to evaluate the performance of the machine, and it is also the loss function of the machines i.e. the value the MLP minimizes to learn the best fit. It is calculated as

\begin{equation}
    \text{MSE}=\frac{1}{n} \sum_{i=1}^{n}(y_i - \hat{y}_i)^2 \text{ ,}
\end{equation}
where $n$ is the size of the sample, $y_i$ is the real value, and $\hat{y}_i$ is the predicted value.

This metric evaluates how far the predictions are from the real values. Therefore, better predictions will produce a MSE  closer to 0.

\subsubsection{Coefficient of Determination ($\text{R}^2$)}

The coefficient of determination, or $\text{R}^2$ score, represents the proportion of variance in the predicted values that can be explained from the observed values. Instead of correlation between variables, it explains to what extent the variance of the real values explain the variance of the predicted values. To calculate $\text{R}^2$ we must take in account the residual sum of squares (RSS) and the total sum of squares (TSS). These values are computed as:

\begin{equation}
    \text{RSS} = \sum_{i=1}^{n}(y_i - \hat{y}_i)^2
\end{equation}

\begin{equation}
    \text{TSS} = \sum_{i=1}^{n}(y_i - \bar{y}_i)^2\text{ ,}
\end{equation}
where $n$ is the size of the sample, $y_i$ is the real value, $\hat{y}_i$ is the predicted value, and $\bar{y}_i$ is the mean of the real values.

Finally, $\text{R}^2$ is calculated as:

\begin{equation}
    \text{R}^2= 1-\frac{\text{RSS}}{\text{TSS}}.
\end{equation}
As opposed to MSE, $\text{R}^2$ will be higher (and closer to 1) for good predictions.

\subsubsection{Pearson Correlation Coefficient}

The Pearson Correlation Coefficient (PCC) measures linear correlation between two predicted and real values. It is the ratio between the covariance of two variables and the product of their standard deviations. It is a normalised measurement of the covariance, with values that range between -1 and 1. PCC is calculated as:

\begin{equation}
    \text{PCC} = \frac{\text{cov}(y,\hat{y})}{\sigma_{y}\sigma_{\hat{y}}}
\end{equation}
where $\text{cov}(y,\hat{y})$ is the covariance between real and predicted values, $\sigma_{y}$ is the standard deviation of the real values, and $\sigma_{\hat{y}}$ is the standard deviation of the predicted values

For PCC, a value of 1 indicates a perfect linear correlation and a value of -1 indicates a perfect inverse correlation. A PCC of 0 indicates no correlation whatsoever. Since we seek to match exact values, we aim for a PCC as close to 1 as possible.



\subsection{Chained-network method: SFR+} \label{sec:chain}

%
As previously discussed in section \ref{sec:theory}, while SFR shows relation with the  dark matter halo properties, stellar mass has a very significant relation with SFR in galaxies. While we cannot measure stellar mass from a DM-Only simulation, we can estimate it and use it as input, as previously done by \cite{Jo_2019}. In this case, and based on the quality of the stellar mass predictions presented in section \ref{chap:results} and discussed in \ref{chap:discussion}, we use the pre-trained MLP described in section \ref{sub:setup}, trained using a dataset of 42705 galaxies, to predict the stellar mass of a galaxy and then use this prediction as another input to train a machine able to predict SFR (and meanSFR). The idea of chaining of different estimators to improve results was first addressed by \cite{WOLPERT1992241} and by \citep{Ting_1999}, where is referred as Stacked Generalization. This ensemble method is not restricted just to neural networks, but the concept is tightly related with residual neural networks \citep{he2015deep}. When presenting results and in the discussion, we will address machines using chained networks as SFR+ and meanSFR+, i.e., adding the plus sign to the shorthand.

\section{Results}\label{chap:results}

\subsection{Stellar mass} \label{sec:stresults}

\begin{figure}
    \centering
    \includegraphics[width = 0.35\textwidth]{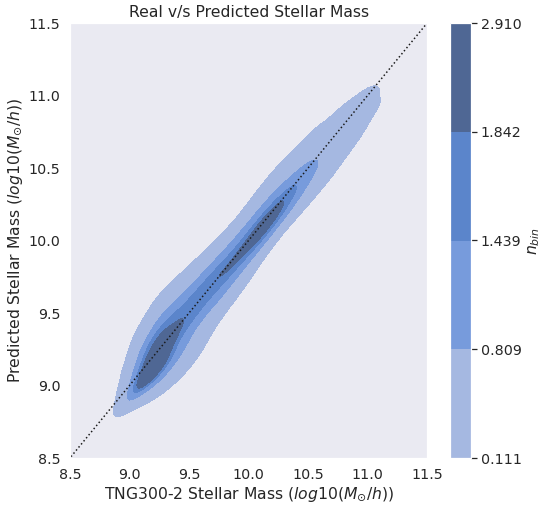}
    \caption{KDE Plot comparing the actual stellar masses from the TNG300-2, and the stellar masses predicted by the MLP trained with the TNG300-2 DM-Only data. Darker shades of blue indicate a higher density of points. $n_{bin}$ represents the number of dots per bin. Both sets of stellar masses are divided in 200 bins. The black dotted line represents an ideal prediction.}
    \label{fig:kdestellar}
\end{figure}

\begin{figure}
    \centering
    \includegraphics[width = 0.35\textwidth]{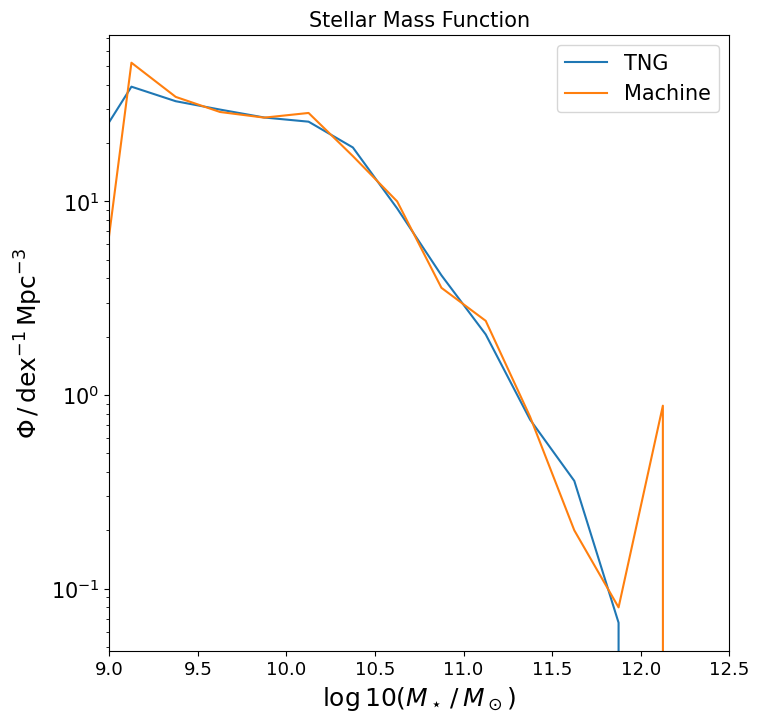}
    \caption{Stellar Mass Function (SMF) of predicted and TNG values. The predicted SMF closely resembles the TNG SMF between 9 and 12 $log_{10}(m_{\odot})$, but it has a sharp increase at high masses, likely due to small number statistics. This behaviour can be seen in figure \ref{fig:weirdupperbound} and is discussed in section \ref{sec:stresults}.}
    \label{fig:stellarmassfunction}
\end{figure}

\begin{figure}
    \centering
    \includegraphics[width = 0.47\textwidth]{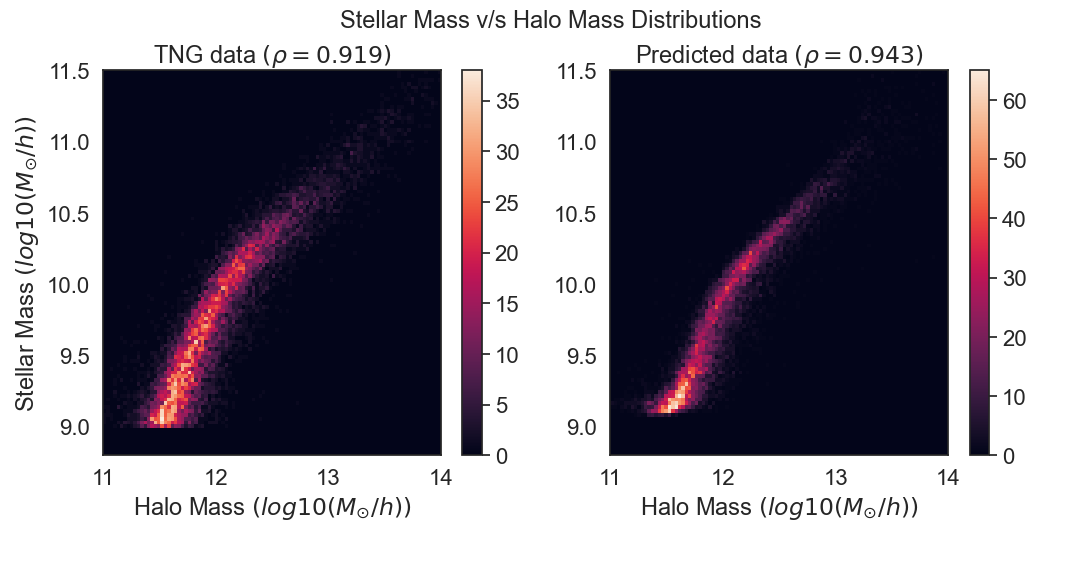}
    \caption{Density plots of the TNG and predicted stellar mass v/s Halo Mass for the sample. Colors indicate the number of samples in each bin. The comparison between both plots hints that the scatter and distribution are slightly off from the simulated data. There appears to be an overdensity on lower values in the predicted values (the colorbar of the right figure reach values, which represent density, higher than 60 points per pixel, while the left figure only reaches about 40 points per pixel). We can also see this density difference in  at lower values in the KDE plot in figure \ref{fig:kdecomparisonStmass}}
    \label{fig:stmassscatter}
\end{figure}

\begin{figure}
    \centering
    \includegraphics[width = 0.45\textwidth]{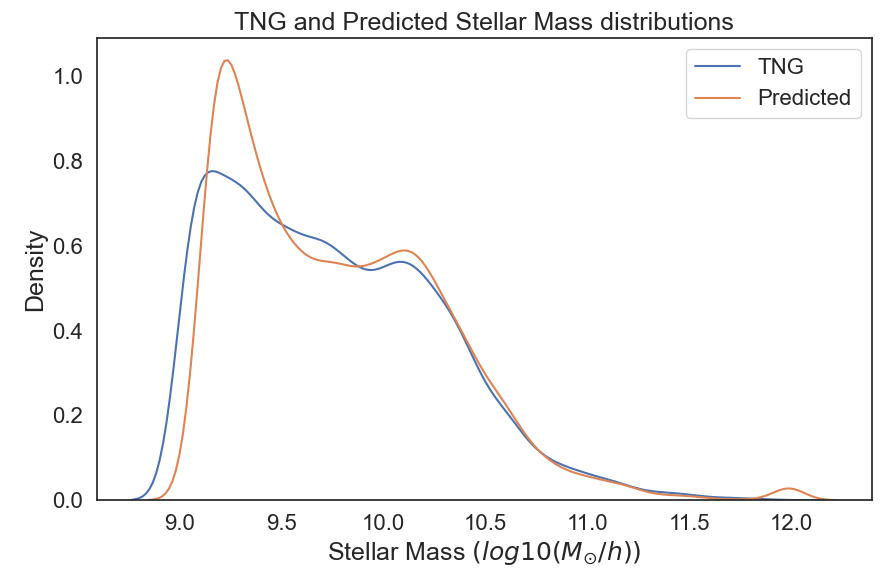}
    \caption{Distributions of predicted and TNG Stellar Mass values. We can see that the predicted distribution presents a higher density at lower values, which is consistent with the overdensity observed in figure \ref{fig:stmassscatter}.}
    \label{fig:kdecomparisonStmass}
\end{figure}

\begin{figure}
    \centering
    \includegraphics[width = 0.45\textwidth]{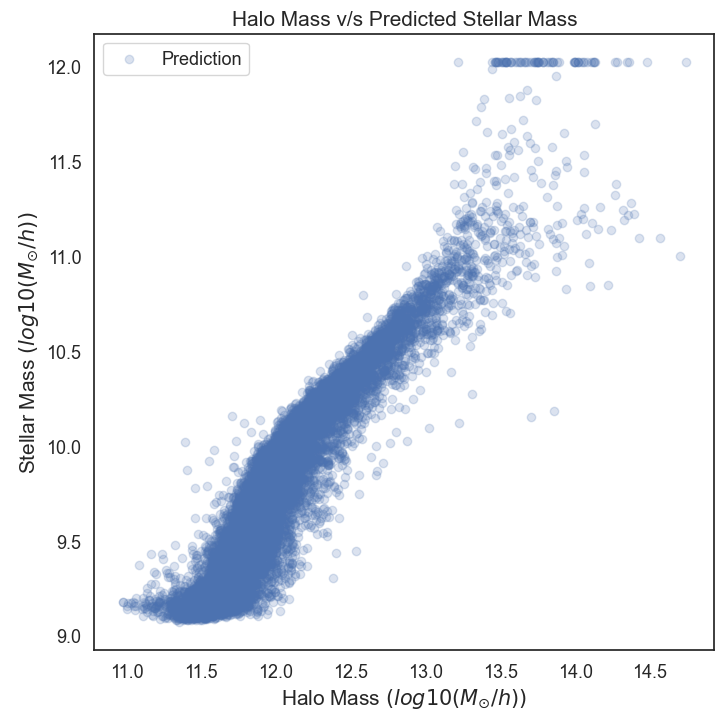}
    \caption{Scatter plot showing Halo Mass against predicted Stellar Mass values. We can see there are some values that are mapped to a constant upper bound, which appears as a horizontal line in the higher values of the scatter plot. The percentage of values presenting this behaviour only corresponds to 0.6\% of the predictions.}
    \label{fig:weirdupperbound}
\end{figure}

After applying the constrains mentioned in section \ref{sec:constraints} our dataset contains a total of 71177 galaxies, divided into 42705 galaxies for the training set and 14236 for both validation and testing set. The metrics for the predictions made by the machine with this training are presented in Table \ref{table:mainresults} and the KDE plot of the TNG and predicted values is presented in Figure \ref{fig:kdestellar}. The KDE plot shows that the results tightly agglomerate around the ideal prediction. This indicates that the predicted stellar masses closely resemble the real values. We can also see in Figure \ref{fig:stellarmassfunction} that the stellar mass function (SMF) of the predicted values closely resembles the original TNG SMF, indicating that the distribution of stellar masses is recovered by the machine, but presents a peak on higher values. The distribution also presents a distinct peak on lower values, as shown in figures \ref{fig:stmassscatter} and  \ref{fig:weirdupperbound}. Figure \ref{fig:stmassscatter} shows the distribution of predicted and real values for Stellar and Halo masses. The difference in PCC implies that there is a stronger linear correlation between these variables in the prediction than in the TNG data. This hints that the scatter and distribution of the prediction are slightly off from the simulated data. To quantify the difference in the scatter we compute the distance correlation between halo mass and stellar mass of TNG and predicted values. This measure is closer to 1 for variables that are less scattered from their mean relation. In the case of TNG values, we get a distance correlation of 0.908 and in the prediction, we get a distance correlation of 0.935, which implies a higher scatter in the TNG values in the stellar mass to halo mass relation. On top of that, we can see that the machine is unable to predict the lower values of the real sample. We also notice what, for high halo masses, there appears to be a constant upper threshold in the predictions from the machine, as shown in figure \ref{fig:kdecomparisonStmass}. This phenomenon can also be seen at the higher mass end in Figure \ref{fig:stellarmassfunction} and Figure \ref{fig:weirdupperbound}. As these figures show, this behavior only occurs at high halo masses, where there is only a small number of galaxies, only 0.6\%  of the total sample. With this in mind, we infer that this is due to small number statistics. Also, in the same figure, a sharp drop at lower stellar masses can be seen, which is consistent with the inability to reproduce lower values previously mentioned. Finally, the metrics of our results show that our machine outperforms the one trained in \cite{Agarwal_2018}, presented in Table \ref{table:agarwal}. They report better results from a random forest than from a neural network. This suggests that a proper hyperparameter tuning, plus the new features we introduced, make neural networks a stronger candidate for performing predictions from dark matter simulations.

\begin{table}
    \centering
    \begin{tabular}{|c|c|c|c|c|c|}
    \hline
          & Stellar mass & SFR & SFR+ & meanSFR &  meanSFR+\\
         \hline
        MSE & 0.017 & 0.147 & 0.144 & 0.140 & 0.135\\
        $R^2$ & 0.937 & 0.300 & 0.378 & 0.517 & 0.576\\
        PCC & 0.970 & 0.776 & 0.780 & 0.828 & 0.833\\
        \hline
    \end{tabular}
    \caption{Final metrics of the predictions in the validation} 
    \label{table:mainresults}
\end{table}

\begin{figure}
    \centering
    \includegraphics[width = 0.45\textwidth]{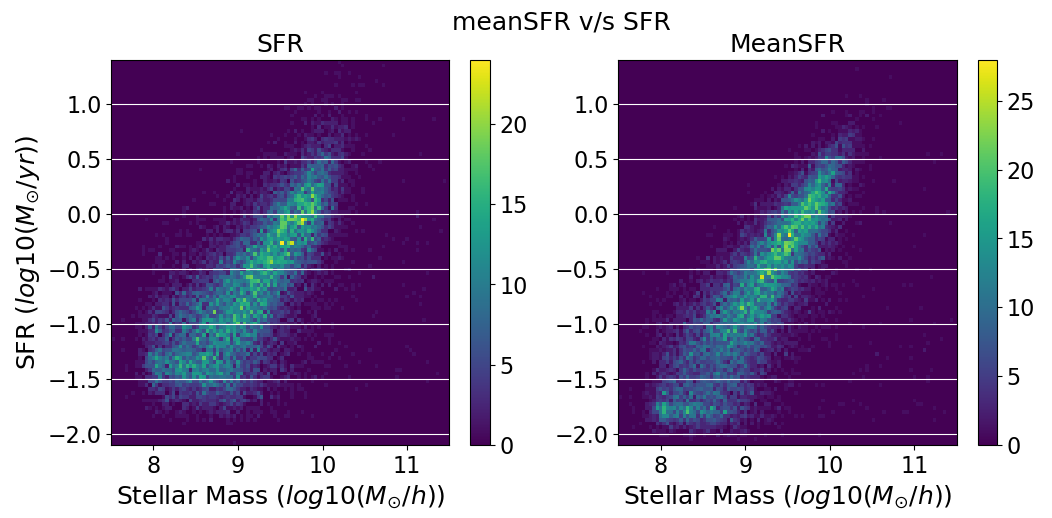}
    \caption{2D histogram comparing SFR and meanSFR distributions with respect to stellar mass. The latter is a good predictor for SFR as discussed in Section \ref{sec:theory}. We can appreciate a higher scatter in SFR than in meanSFR around the main body of values. On top of that, the distribution of meanSFR propagates to lower values, as the horizontal grid lines allow to observe. This is because some galaxies are not star forming at $z=0.1$, and therefore the SFR is reduced when calculating the mean between redshifts.}
    \label{fig:sfrscatter}
\end{figure}

\begin{figure}
    \centering
    \includegraphics[width = 0.45\textwidth]{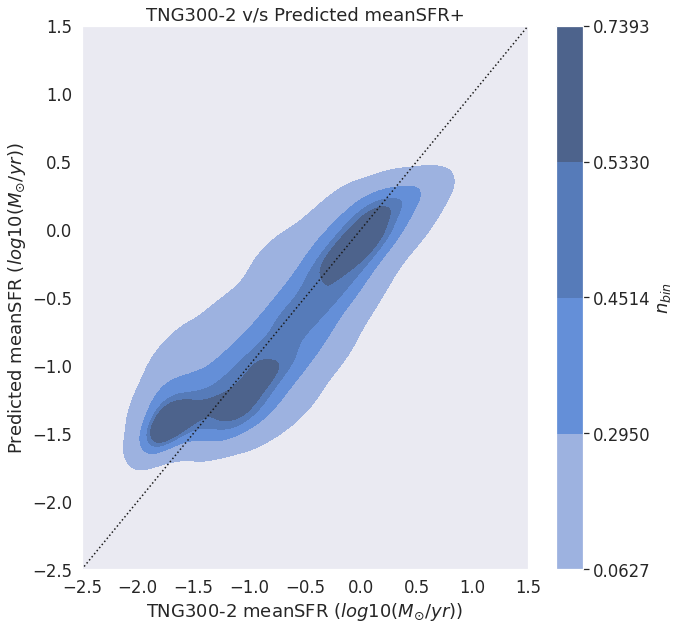}
    \caption{KDE Plot comparing meanSFR computed from the TNG300-2, and the meanSFR predicted by the MLP trained with the TNG300-2 DM-Only data with the meanSFR+ method. Darker shades of blue indicate a higher density of points. $n_{bin}$ represents the number of objects in one bin. Both sets of SFRs are divided in 200 bins by default. The black, dotted line represents the ideal prediction. The density being above the ideal prediction at lower values and under the same line at higher values shows that the machine is not predicting values in the same range as the TNG data.}
    \label{fig:sfrkde}
\end{figure}

\begin{figure}
    \centering
    \includegraphics[width = 0.45\textwidth]{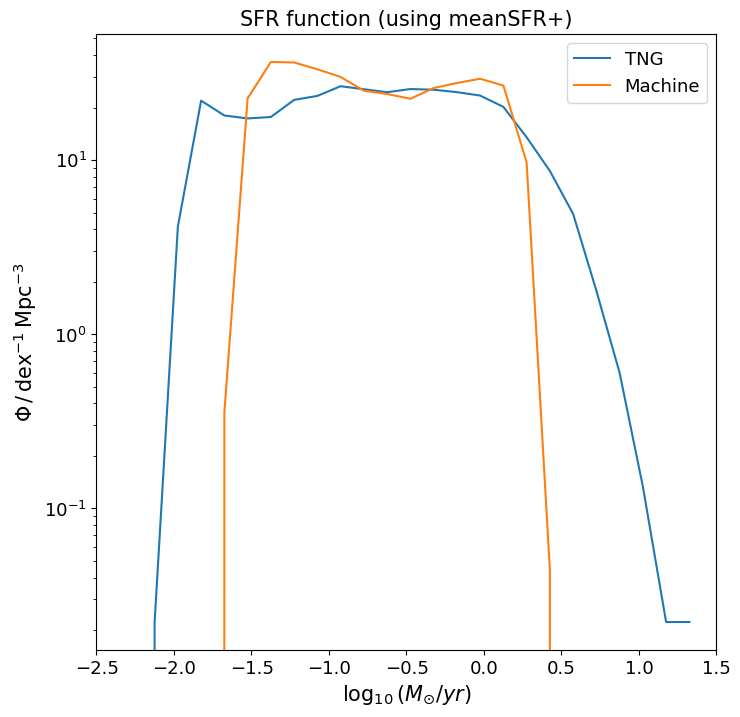}
    \caption{Star formation rate Function of TNG and  machine predicted values (different colours, shown in the figure key). We use meanSFR as the TNG value and the meanSFR+ for the predicted value. We can see the machine overpredicts low SFR values and cannot predict values past a threshold, as discussed in section \ref{subsec:msfrp}}
    \label{fig:sfrf}
\end{figure}

\begin{figure}
    \centering
    \includegraphics[width = 0.5\textwidth]{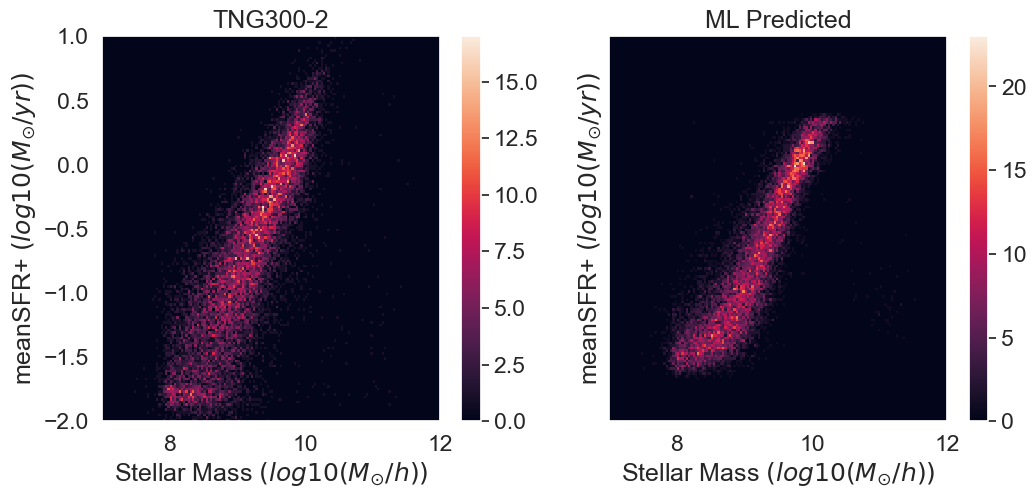}
    \caption{Scatter plots comparing the distributions of TNG and predicted meanSFR+ values vs. stellar mass. In the left we have the distribution of TNG values and in the right the predicted values. We can see the machine having problems at reproducing the natural scatter of this property, and its inability to predict high and low values of mean star formation.} 
    \label{fig:sfrdist}
\end{figure}

\begin{figure*}
    \centering
    \includegraphics[width = 0.75\textwidth]{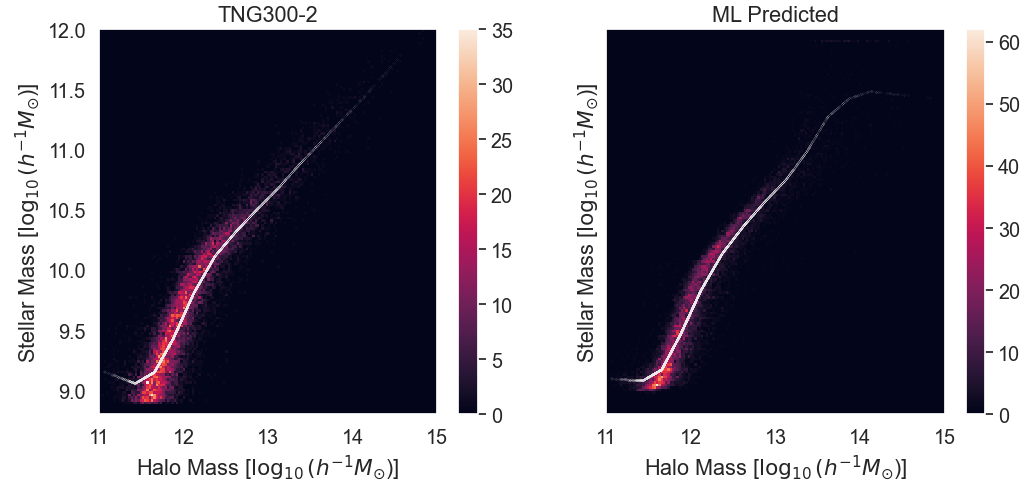}
    \caption{Scatter plots showing the distribution of stellar mass in relation to halo mass. In the left figure, we can see TNG data. In the right figure, we see the prediction's distribution. We can picture an erratic behaviour at high halo masses for the prediction due to low number statistics.} 
    \label{fig:deltascatterstar}
\end{figure*}

\begin{figure}
    \centering
    \includegraphics[width = 0.45\textwidth]{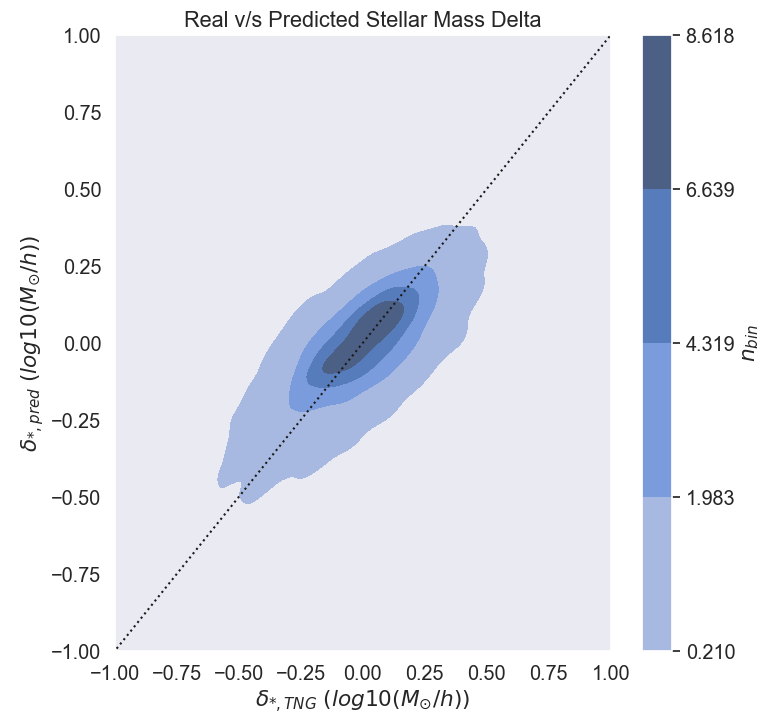}
    \caption{KDE plot comparing the difference from the mean value of the Halo Mass/stellar mass relation. We can see that the machine is able to partially reproduce the sign of the deviation from the mean relation present in the TNG galaxies.} 
    \label{fig:deltastar}
\end{figure}

\begin{figure*}
    \centering
    \includegraphics[width = 0.75\textwidth]{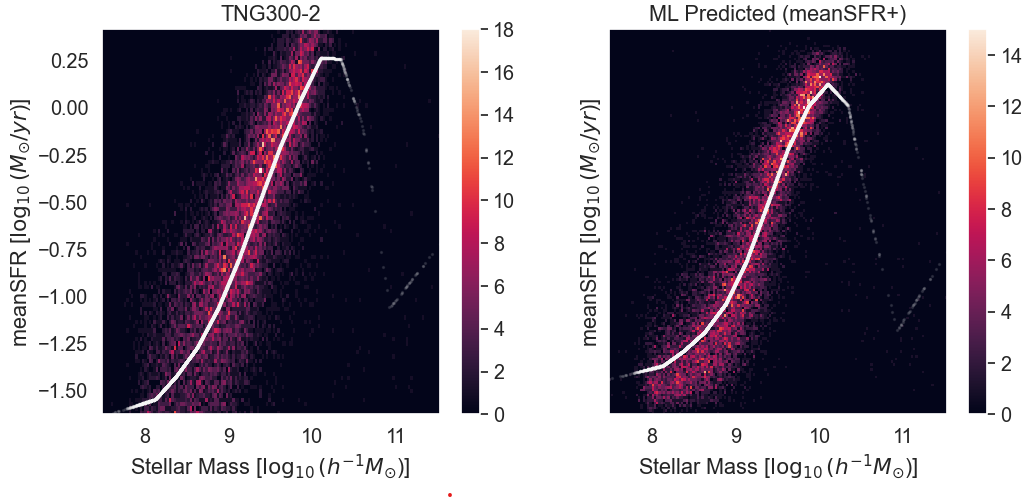}
    \caption{Scatter plots showing the distribution of star formation rate in relation to stellar mass. In the left figure, we can see TNG data. In the right figure, we see the prediction's distribution. The white dotted line represents a piecewise function, where we bin meanSFR in 20 Stellar Mass bins, and compute the mean of the values for each bin.}
    \label{fig:deltascattersfr}
\end{figure*}

\begin{figure}
    \centering
    \includegraphics[width = 0.45\textwidth]{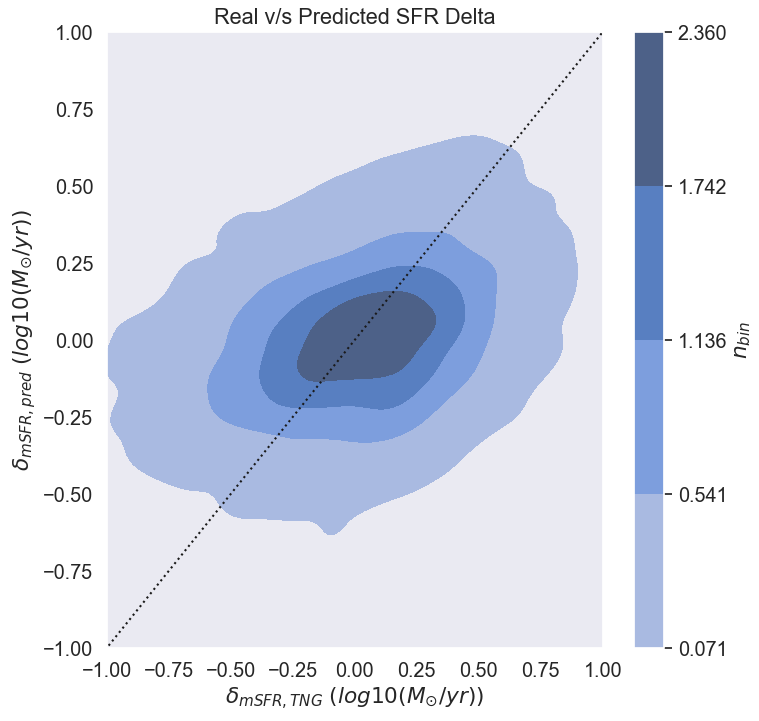}
    \caption{KDE plot comparing the difference from the mean value of the stellar mass/SFR relation. The distribution slightly tilts towards the direction of the identity line, which indicates a weak relation. The range of the machine deltas is narrower. This makes sense since the machine's SFR distribution has a considerably lower scatter.}
    \label{fig:deltasfr}
\end{figure}

\subsection{Star formation rate} \label{sfrres}

SFR proved to be a tricky variable to predict in \cite{Jo_2019} and \cite{Agarwal_2018}. Also, as stated in Section \ref{sec:contrast}, the SFR reported in the Illustris catalog is instantaneous and stochastic. Because of this, we will use ML methods to predict not only SFR at $z=0$, but also the mean of SFRs for the same galaxy between $z=0.1$ and $z=0$. This variable will be addressed as \textit{meanSFR}. This not only allows us to have a variable that can be better compared to SFR measured from observations, which depending on the observed emission lines can respond to different time averaging (for a discussion of the different timescales associated to different observational estimates of the SFR, see for instance \citealt{Guaita2011}), but this choice also reduces the scatter of the variable which cannot be learnt by ML, as seen in figure \ref{fig:sfrscatter}. The latter is convenient for this model since the inability of reproducing the scatter reduces its impact on the results. On top of that, as mentioned in \ref{sec:chain}, we will evaluate the performance of the machine adding an extra variable, the stellar mass predicted by a pre-trained machine on the same dataset. We use this approach on both SFR and meanSFR. When treating these variables with the new input, they will be addressed as "SFR+" and "meanSFR+" respectively.

For all the following trainings, after applying the cuts mentioned in section \ref{sec:constraints}, our dataset contains 81810 galaxies, divided in 47298 for the training set and 15767 for both validation and testing set.

\subsubsection{Metrics for predictions of SFR} \label{subsec:sfr}

The metrics made by the machine with this training are presented in Table \ref{table:mainresults}. As shown in this table, this is the worst of the 4 regressing methods for SFR. Nontheless, the PCC metric surpasses the values obtained by \cite{Agarwal_2018} shown in Table \ref{table:agarwal}, but our $\text{R}^2$ score is considerably lower for the same prediction. As discussed in Section \ref{subsec:metrics}, this means that, while the predicted and real values are more correlated, the scatter in the predicted values does not resemble the scatter in the TNG values, which is better reproduced in \cite{Agarwal_2018}.

\subsubsection{Metrics for predictions of SFR+}

The metrics made by the machine with this training are presented in Table \ref{table:mainresults}. We can see an improvement in all metrics when compared with SFR. While the results are better, we still find that only the PCC metric is better than the one at \cite{Agarwal_2018} shown in Table \ref{table:agarwal}, while $\text{R}^2$ score is still lower than theirs. From this comparison in metrics, we can infer the same as in Section \ref{subsec:sfr} about how the data correlates.

\subsubsection{Metrics for predictions of MeanSFR}

We can see an important improvement in the metrics in this case. This method improves the metrics over SFR better than the SFR+ method. The $\text{R}^2$ score metric is the one that sees the higher improvement. This makes sense when considering the discussion in Section \ref{sec:theory} and the scatter plot in Figure \ref{fig:sfrscatter}, since a lower scatter in values makes $\text{R}^2$ a more forgiving metric.

\subsubsection{Metrics for predictions of MeanSFR+} \label{subsec:msfrp}

This approach, combining both previous methods, gives the best results obtained for SFR regressions. This time, both $\text{R}^2$ and PCC are better than the ones from \cite{Agarwal_2018}. For this regression, the KDE plot is presented in Figure \ref{fig:sfrkde}, which compares the meanSFR values of TNG against predicted values using the meanSFR+ method. It can be seen that while the distribution lies around the ideal prediction, the machine overpredicts values for low SFR and underpredicts values for high SFR. This can be seen in the density of predictions lying over the ideal prediction line for lower meanSFR values, and the density lying under the ideal prediction at higher values. This is consistent with the scatters observed in Figure \ref{fig:sfrdist} and with the SFR function in Figure \ref{fig:sfrf},  where the machine is seen to be unable to reproduce the lower and higher values of the mean SFR from TNG using the meanSFR+ method. 

\subsection{Predicting departures from mean values and their sign.} \label{sec:deltas}

As stated in section \ref{sec:theory}, observations indicate that a mean relation between stellar mass and halo mass, and also between SFR and stellar mass, can be estimated from observations. With this in mind, we will explore how well our machine predicts in terms on how far each galaxy is from the mean relation. In other words, we are interested in studying if the deviation from the mean relation (to either higher or lower values) is recovered. We will study how well can we recover this deviation, or \textit{delta}, for stellar mass and meanSFR. We calculate this delta from mean relations as:

\begin{equation} \label{delta1}
    \delta_{M^*}= M_{*,TNG}- \langle M_{*,TNG} \rangle_{M_{halo,TNG}}
\end{equation}

\begin{equation} \label{delta2}
    \delta_{mSFR}= \text{mSFR}_{TNG}- \langle \text{mSFR}_{TNG} \rangle_{M_{*,TNG}}\text{ ,}
\end{equation}

where $M_{*,TNG}$ is the stellar mass from TNG300-2, $\langle m_{*,TNG} \rangle_{M_{halo,TNG}}$ is the value at the same halo mass from the mean relation of TNG's stellar mass with respect to TNG's DM-only halo mass, $\text{mSFR}_{TNG}$ is the meanSFR from TNG300-2, and $\langle \text{mSFR}_{TNG} \rangle_{M_{*,TNG}}$ is the value at the same stellar mass from the mean relation of TNG's meanSFR with respect to TNG's stellar mass.

In both equations, we compute the mean relation as a piecewise linear function that goes through the mean values of binned stellar masses (or meanSFRs), in bin intervals of halo mass (or stellar mass).

For calculating predicted values' means, we change the predicted value for its respective TNG value.

\subsubsection{Stellar mass}

In the case of stellar mass, the mean relation is well predicted for $log(m_{halo}) < 13m_{\odot}$. After that, the scatter plus the upper threshold predictions mentioned in section \ref{sec:stresults} make the mean relation to deviate upwards. With that being said, we can see in the KDE plot in figure \ref{fig:deltastar} that there is a correlation between the deltas from TNG and the simulation. While the deltas are not tightly gathered around the ideal prediction, we can see the orientation of the distribution follows the identity line, which means the sign of the deviation from the mean is well predicted.

\subsubsection{MeanSFR} 

For this variable, the lack of values at upper and lower values makes the mean relation to have a narrower curve, as seen in figure \ref{fig:deltascattersfr}. With this said, it reproduces the quenching of galaxies at high masses discussed at section \ref{sec:contrast}, and the sudden increase for $M^* > 11*10^{10}m_{\odot}$. Since the prediction presents less scatter, it is reasonable to assume that the deltas in this case will be smaller. In figure \ref{fig:deltasfr} we can see this is the case, but the correlation between the sign of the deviations seem weakly correlated.

\subsection{Influence of input variables in results}

\subsubsection{Parameter ranking} \label{sub:rank}

From comparisons with previous work and by studying the chained-network method, we observe that adding new variables impacts positively the performance of the machine. The parameter ranking has been approached in different ways in studies similar to this one. In \cite{Jo_2019} and \cite{Agarwal_2018}, where they use random forests to make the regression, they count how many times a variable appears in the trees, with the most important features appearing more times.  \cite{Calderon_2019} use the same approach and also use an xgboost regressor which natively computes feature importance to study the impact of each input variable.  \cite{Shao_2022}, on the other hand, use saliency values which identify the most important variables that contribute to the relationship between all inputs and outputs.

In this work we take a different approach 
for checking how each input variable affects the model 
running multiple trainings removing one input feature at a time. We run 10 trainings for each variable, to check that the change in the performance is smaller than the standard deviation of the metrics, allowing to check if the difference in the performance is influenced by the removed variable or by the intrinsic stochasticity of NN. Because of their similar nature, we removed $m_{halo}$ and $m_{crit,200}$ together, because one variable could give information contained in the other one. The same applies for $r_{crit,200}$ and $r_{crit,500}$.

\begin{figure}
    \centering
    \includegraphics[width = 0.5\textwidth]{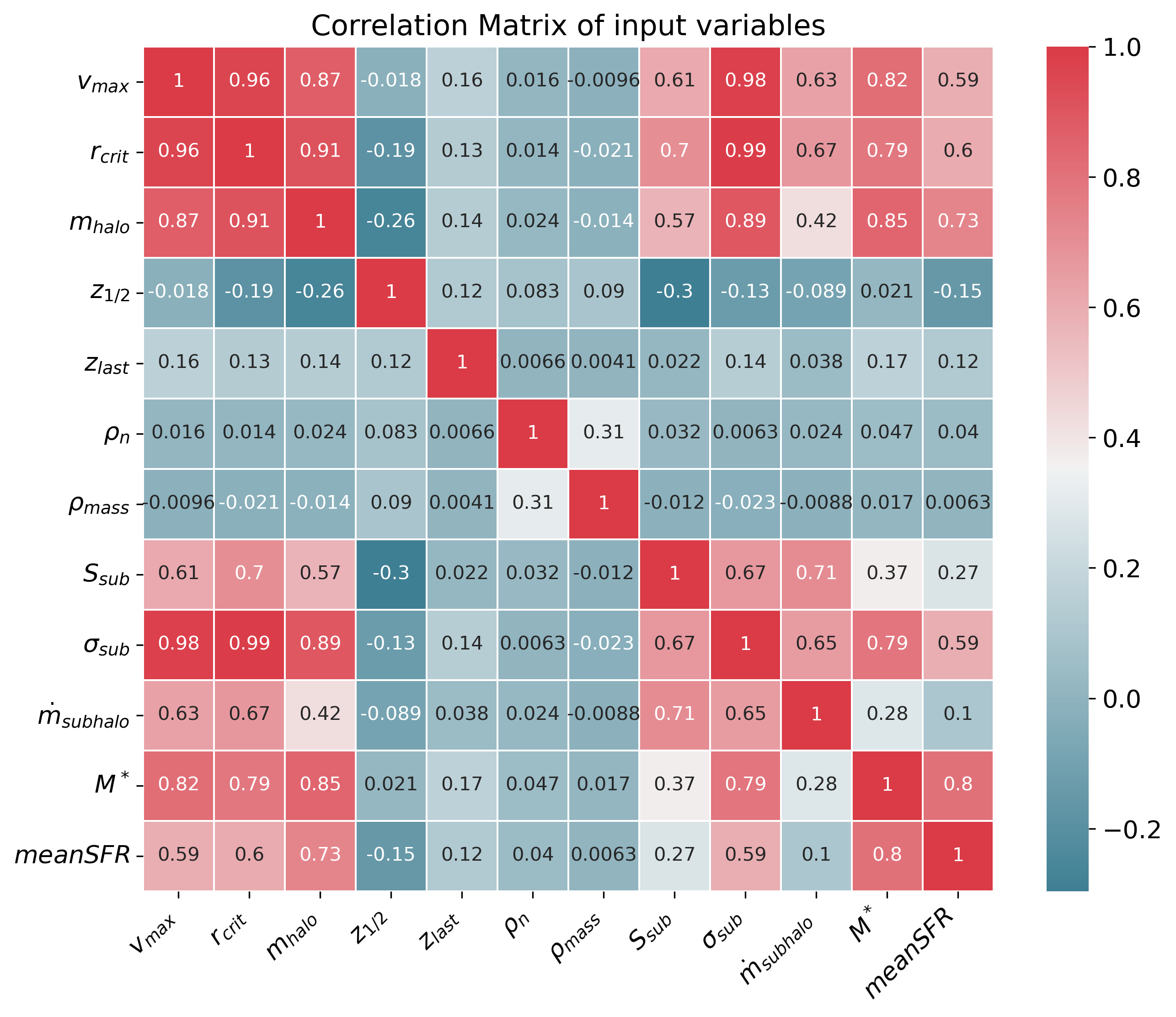}
    \caption{Correlation matrix between input and output variables. We can see a particularly high correlation between input variables $v_{max}$, $m_{halo}$, $r_{crit}$ and $\sigma_\text{sub}$. As \ref{table:stmassrank} and \ref{table:sfrrank} show, $m_{halo}$ and $\sigma_\text{sub}$ doesn't impact the model when removed by themselves, which we infer is due to the high correlation to the more impactful variables $v_{max}$ and $r_{crit}$. We can also observe that both output variables, $M^*$ and $meanSFR$ have higher correlation with those four input variables.}
    \label{fig:corr_x}
\end{figure}

Table \ref{table:stmassrank} shows that, for $M^*$, the maximum circular velocity was the most impactful feature to remove, followed by the $r_{\rm crit}$. The maximum velocity being a good predictor of the stellar mass is a result that has been reported already \citep{vmaxhalo}, which supports our results. In third and fourth place we have $z_{1/2}$ and $z_{last}$. 
This suggests that the history of the evolution of the dark matter medium has a strong influence on this variable, not only in terms of the number of mergers or their mass ratio as studied in \cite{Jo_2019}, but also the age of the universe for these events. In the case of halo mass, it is not very intuitive for it to come in fifth place. While the halo mass's strong correlation with stellar mass has been thoroughly investigated, one may argue that $v_{max}$ and $r_{crit}$ are strongly correlated with $m_{halo}$ and therefore, the machine can infer to some extent this relation with halo mass \citep{vmaxhalo}.

Figure \ref{fig:corr_x} presents a correlation matrix between all input and output variables, with the goal to study the influence between variables and gain insights into how the information they provide, or the absence of it, impacts the model. This matrix shows that the three aforementioned input variables are strongly correlated. The ranking continues with the environmental properties. From the table, we can see these variables improve the results of the predictions, but they are clearly not as influential as historical or environmental dark matter properties. Finally, we highlight the low influence shown by the halo spin and velocity dispersion. These two variables play an important role in \cite{Jo_2019} and \cite{Agarwal_2018}, where they are relatively well predicted. We believe that this is relevant enough to be studied on its own, and while we conjecture that the relevant information these variables contribute is encompassed by  other variables, deepening these relations is beyond the scope of this work.

On the other hand, for SFR Table \ref{table:sfrrank} shows the predicted stellar mass as the most impactful feature when removed. This reinforces the value of adding such a feature to the input data and the consistency of chaining networks. Once again, $z_{1/2}$ appears high in the ranking, showing the importance of historical data for inferring baryonic properties. While $m_{halo}$ coming in third place is relevant to mention, the fourth place deserves special attention. $\dot{m}_{subhalo}$ is another variable that was not explicitly seen in literature, and it suggests a relation between the accretion of a subhalo and the star formation happening inside. In fifth place, and very close to $\dot{m}_{subhalo}$ comes $z_{last}$, once again showing the importance of historical properties.  Environmental properties have a positive, but smaller impact. We interpret this as them being consistently influential in the studied galaxy properties, even if said influence is overshadowed by more impactful variables. The spin and velocity dispersion have again a low place in the ranking. But this time, unlike with stellar mass, the critical radius and maximum circular velocity are also low in the ranking. While this may reinforce the conjecture made earlier in this section, we infer that this may also suggest that the dynamic properties of a halo are not as important as historical and environmental properties on the SFR, though to study this in depth is beyond the scope of this work. We deepen the discussion about input variables in section \ref{remarkin}.

\begin{table*}[b]
    \centering
    \begin{tabular}{|c|c|c|c|c|c|c|}
    \hline
    Feature &    MSE &  MSE \% &  $\text{R}^2$ Score &  $\text{R}^2$ \% &  PCC &  PCC \% \\
    \hline
          $v_{max}$  &  0.0194 &   14.378\% &    0.9234 &   0.467\% &        0.9655 &    1.456\% \\
            $r_{crit}$  &  0.0190 &   11.614\% &    0.9253 &   0.329\% &        0.9668 &    1.253\% \\
       $z_{1/2}$ &  0.0184 &    8.226\% &    0.9288 &   0.274\% &        0.9673 &    0.872\% \\
          $z_{last}$ &  0.0183 &    7.676\% &    0.9296 &   0.277\% &        0.9673 &    0.791\% \\
      $m_{halo}$ &  0.0179 &    5.572\% &    0.9303 &   0.237\% &        0.9677 &    0.710\% \\
     $\rho_{mass}$ &  0.0179 &    5.434\% &    0.9305 &   0.210\% &        0.9680 &    0.693\% \\
          $\rho_n$ &  0.0177 &    4.252\% &    0.9318 &   0.224\% &        0.9678 &    0.550\% \\
       $S_\text{sub}$ &  0.0177 &    4.089\% &    0.9314 &   0.202\% &        0.9680 &    0.601\% \\
        $\sigma_\text{sub}$ &  0.0177 &    3.981\% &    0.9313 &   0.209\% &        0.9680 &    0.604\% \\
      $\dot{m}_{subhalo}$  &  0.0175 &    3.058\% &    0.9322 &   0.197\% &        0.9681 &    0.516\% \\
    \hline
    \end{tabular}
    \caption{Parameter ranking for stellar mass prediction. Percentages indicate the level to which the metric worsened by removing the corresponding feature in that row, with respect to the best results obtained.} 
    \label{table:stmassrank}

\end{table*}

\begin{table*}[b]
    \centering
    \begin{tabular}{|c|c|c|c|c|c|c|}
    \hline
    Feature &    MSE &  MSE \% &  $\text{R}^2$ Score &  $\text{R}^2$ \% &  PCC &  PCC \% \\
    \hline
    $m_{*,pred}$ & 0.1518 &   12.436\% &    0.4576 &   2.595\% &        0.8114 &   20.558\% \\
     $z_{1/2}$ & 0.1460 &    8.116\% &    0.4989 &   1.647\% &        0.8193 &   13.385\% \\
    $m_{halo}$ & 0.1447 &    7.182\% &    0.5060 &   1.407\% &        0.8213 &   12.150\% \\
     $\dot{m}_{subhalo}$ & 0.1445\% &    7.016 &    0.5078\% &   1.394 &        0.8214 &   11.839\% \\
    $z_{last}$ & 0.1444 &    6.951\% &    0.5082 &   1.392\% &        0.8214 &   11.775\% \\
    $\rho_n$ & 0.1432 &    6.052\% &    0.5103 &   1.166\% &        0.8233 &   11.399\% \\
    $\rho_{mass}$ & 0.1430 &    5.903\% &    0.5116 &   1.128\% &        0.8236 &   11.177\% \\
    $v_{max}$ & 0.1426 &    5.595\% &    0.5174 &   1.081\% &        0.8240 &   10.166\% \\
    $\sigma_\text{sub}$ & 0.1400\% &    3.737 &    0.5377\% &   0.622 &        0.8278 &    6.653\% \\
       $r_{crit}$ & 0.1393\% &    3.156 &    0.5308 &   0.447\% &        0.8293 &    7.849\% \\
    $S_\text{sub}$ & 0.1392\% &    3.082 &    0.5365 &   0.274\% &        0.8307 &    6.864\% \\
    
    \hline
    \end{tabular}
    \caption{Parameter ranking for meanSFR prediction. Percentages indicate how much the metric worsened by removing the corresponding feature in that row, with respect to the best results obtained.} 
    \label{table:sfrrank}

\end{table*}

\subsubsection{Distribution of best and worst prediction inputs} \label{sub:hist}

\begin{figure*}
    \centering
    \includegraphics[width=1.\textwidth]{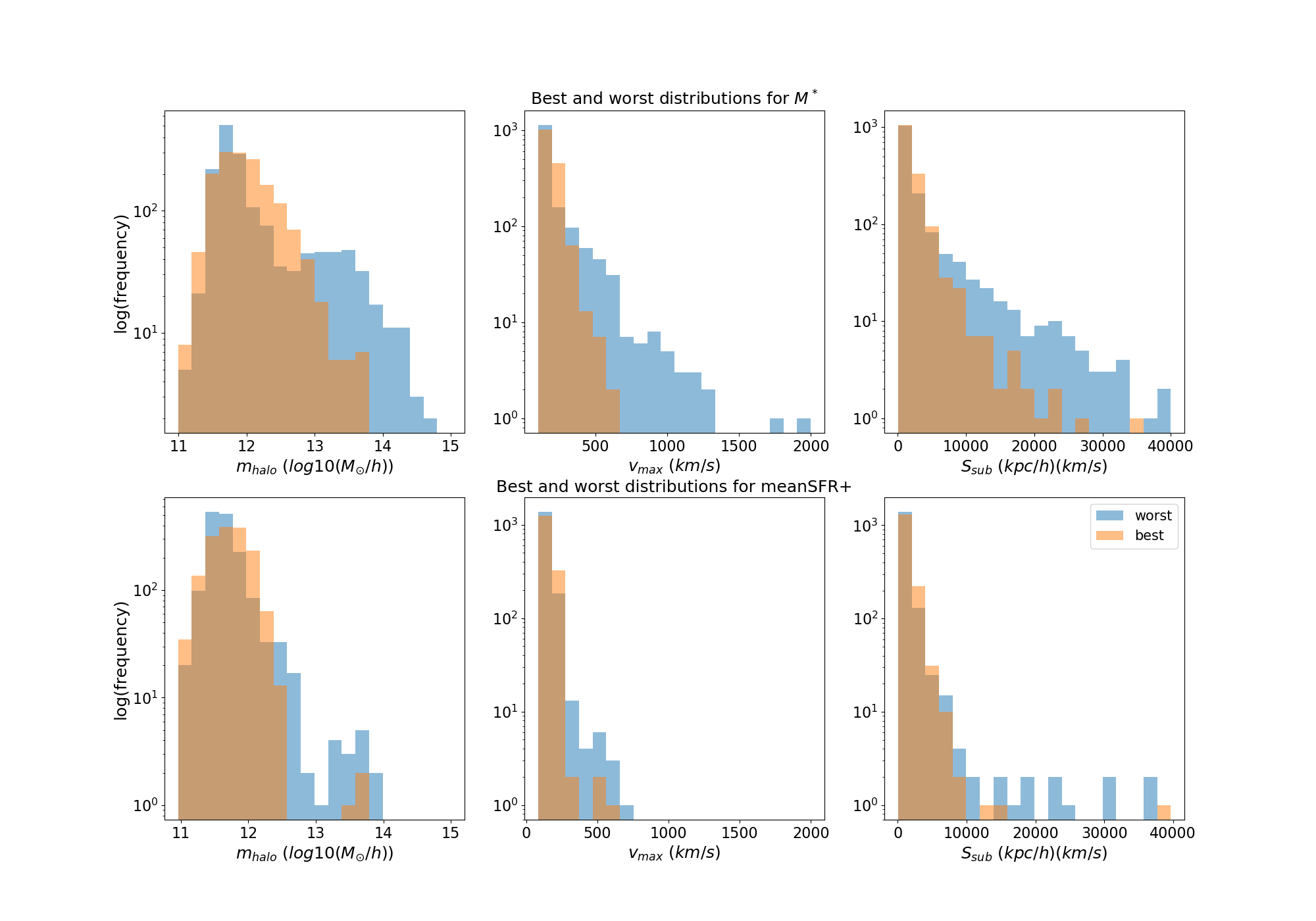}%
    \caption{Histograms of the distribution of $m_{halo}$, $v_{max}$ and $S_{subhalo}$ for the $10\%$ best and worst predictions for $M^*$ and meanSFR. Histograms in the top row belong to $M^*$. Histograms in the bottom row belong to meanSFR. In all histograms, the $y$ axis is log-scaled. The left column shows the $m_{halo}$ distribution, the center column shows the $v_{max}$ distribution, and the right column shows the $S_{subhalo}$ distribution.}
    \label{fig:hists}
\end{figure*}

We will also analyze the histogram distribution of some input variables in the best 10\% and worst 10\% predictions. This will shed light on how these parameters behave and influence the quality of the predictions. These features might not be the most influential overall as seen in Section \ref{sub:rank}, but they present the highest shift between the best and worst predictions. In figure \ref{fig:hists} we can see the frequency distribution for three variables: $M^*$, $v_{max}$ and $S_{subhalo}$. In the case of stellar mass, for the three variables, we can see that the worst predictions tend to have higher values. This trend becomes more evident for the spin and maximum velocity. While not as notorious as in the stellar mass, the three chosen variables also have higher frequencies at higher values for meanSFR. With this on mind, we believe that for future works, this kind of behaviour could be studied in the preprocessing stage. This will be discussed in more detail in section \ref{sec:improve}.

\section{Discussion}\label{chap:discussion}

\subsection{Main Results}\label{sec:mainresults}

In this work, we trained a machine that was able to predict stellar mass in a very consistent way, not only in metrics but also replicating distributions and deviations from mean relations. While the results for meanSFR were not as robust as the ones obtained for stellar mass, our results for both properties still present improved metrics with respect to, for instance, \cite{Agarwal_2018}. We explored approaches to improve the predictions for SFR, where time-averaging SFR and chaining neural networks were successful in doing so. In particular, time-averaging was quite impactful in the results. This approach to predicting SFR and its benefits have not been addressed in similar studies. Time-averaging not only reduced the scatter in the SFR values (which makes it harder for the machine to find an appropriate mapping for), but also makes more sense when taking into account that the SFR measured in real galaxies corresponds to a time averaged quantity. This should be taken into account if a ML method is to be used to reproduce the SFR measurements of particular surveys to match the estimated timescale of averaging present in the observations. This being said, SFR proved to be a tricky property to predict for our model. In section \ref{sec:improve} we discuss ideas for improving predictions.

In terms of the predicted values, both $M^*$ and SFR had issues at the lower and higher ends of the range of values. While some issues can be interpreted from small number statistics, the inability of reproducing extreme values is an important flaw in this method, and should be treated with special attention in future work. The intrinsic scatter of the predicted values could not be fully reproduced, this being particularly true for meanSFR. With respect to the deviation from observed mean relations, the machine was partially successful in doing so for $M^*/m_{halo}$ and, arguably, minimally achieved for meanSFR$/M^*$.

We studied the feature importance and impact of physical properties using a leave-one-covariate-out approach, which has not been previously used in the literature. Doing so, and studying the correlation between input variables, we observe that some variables which would be impactful by themselves (e.g. $m_{halo}$ for $M^*$), when removed do not impact the results of the predictions as much as expected. We infer that this is due to their high correlation to the two most relevant features according to the analysis (e.g. $v_{max}$ and $r_{crit}$ in the case of $M^*$)

Finally, in Section \ref{sub:rank} we discovered that $z_{last}$ and $z_{1/2}$ were quite relevant in the performance of the machine for both $M^*$ and mean SFR. $\dot{m}_{subhalo}$ was also particularly impactful for regressing SFR. These and other interesting features will be discussed in section \ref{remarkin}.

\subsection{Remarkable input variables} \label{remarkin}

As stated in Section \ref{sec:mainresults}, historical properties like $z_{1/2}$ and $z_{last}$ have an important impact on the predictions. This makes sense considering that the properties of a galaxy are a product of its evolution, and shows that this kind of features should be considered as important when generating models to predict baryonic features of galaxies. On top of that, $\dot{m}_{subhalo}$, which could also be seen as a historical property (since it considers data from previous redshifts) had a surprising impact in meanSFR. These variables had not been addressed in similar studies 
(although  \cite{Jo_2019}   used other historical properties).

Other input feature of interest, particularly for $M^*$, is $v_{max}$. Naively one tends to think of the mass of the halo  as the main descriptor of stellar mass, however, it has been shown in the literature that there is a stronger relation with the maximum circular velocity (see for instance \citealt{Kulier}). Therefore it is quite remarkable that the machine is able to identify its importance. 
The information lost by taking $m_{halo}$ from the inputs might be compensated by the critical radius and maximum circular velocity. As figure \ref{fig:corr_x} shows, the correlation between these three variables is very high, which supports this inference. 

Regarding the variables studied in Section \ref{sub:hist}, we can observe in figure \ref{fig:hists} that the top left histogram shows that better $M^*$ predictions tend to have lower halo mass values. This is also true for $v_{max}$ and $S_{subhalo}$ in the upper middle and upper left histograms. For these variables, the inclination towards low values for good predictions is more evident, although these values (unlike $m_{halo}$) are not in logarithmic scale. The same patterns are also present in the three histograms for meanSFR in the bottom row. There also seems to be a tendency to have better results at lower values, but the impact is not as clear as it is for $M^*$. In overall, this shows that bad predictions tend to have higher input values. We suggest that these variables should be treated carefully and taken in special consideration when exploring methods similar to the one presented here.

\subsection{On how can this method be improved} \label{sec:improve}

As shown in the results section, our method struggled to replicate results in the upper and lower limits of the range of predicted properties as seen in Figure \ref{fig:sfrdist}, where the machine is unable to reproduce the range of values; and Figure \ref{fig:stmassscatter}, where the machine could not predict values in the lower bound and mapped some values to the same upper threshold. The limits issue could be approached by exploring other kinds of normalization, such as gaussian or quantile normalizations.

We should also address the stack of values of predicted $M^*$ for large halo masses. This is a direct sign that a larger dataset could improve results, as we can see a decay in the performance at high masses where there is a smaller number of galaxies, although a weighting scheme may be needed for the machine to learn about the sparse sample of objects with the highest masses. We propose tackling this issue by balancing the dataset either generating synthetic data to have more high mass candidates for the machine to learn from, or by reducing the number of galaxies with average values to have a homogeneous distribution of galaxies of different nature. This second approach  might hurt the performance of the machine because of the reduced amount of data. 
Another possibility is to remove  outlier values with clustering methods (like DBSCAN) to reduce the amount of subhaloes with extreme values that might affect the learning of the machine.
Alternatively, in order to improve the performance 
one could try log-scaling the input features or using a principal component analysis before inputting values into the MLP.

Several works in the literature have already shown other methods to improve the tail of the distributions (see, \cite{Jo_2019}; \cite{de_Santi_2022}; \cite{Stiskalek_2022}). For instance, modifying the loss function in \cite{Jo_2019} and, as proposed early for balancing the dataset, it was already done in \cite{de_Santi_2022}. These improvements are beyond the scope of this study, but it will be interesting to address in future research.

As we saw in Sections \ref{sfrres} and \ref{sub:rank}, chaining networks by using a predicted feature as input to predict a second feature remarkably improved the results. With this in mind, the model could be improved by predicting new baryonic properties which, if they can be robustly predicted, may be used as inputs to predict more erratic properties such as the SFR. An example of a feature that could be used this way is metallicity, which was consistently predicted in \cite{Agarwal_2018}.

In section \ref{sub:hyper}, we discussed about how we explored the hyperparameter space to fine-tune our machine. The employed method, where we manually went over different combinations of hyperparameters is called grid search, and it is one of the most basic approaches to this end. In future works, one might employ more sophisticated hyperparameter optimization methods, eg. gradient-based or bayesian optimizations.

For this work we used only 12 input features ($S_\text{sub}$, $\sigma_\text{sub}$, $v_{max}$, $m_{halo}$, $r_{crit,200}$, $r_{crit,500}$, $m_{crit,200}$, $\rho_n$, $\rho_{mass}$, $z_{1/2}$, $z_{last}$, and predicted $m_{*}$ only for meanSFR). Considering the amount of data products that come from a DM-Only discussed in section \ref{sec:product}, the number of input features could be increased. Particularly, we propose that considering more historical features could greatly improve the ability of the machine to predict complex attributes like SFR. Considering how important the halo growth history proved to be, adding data from earlier redshifts might also improve the method.

The MLP is one of the most basic NN architectures. This is not an issue in this work, since we use little input data. But if the amount of data were increased, one could make use of more sophisticated machine learning methods to process said data. If, for example, we decide to use direct historical data (eg. the masses measured in previous snapshots), recurrent neural networks could be an excellent candidate since they have the advantage of being efficient at learning from sequential data \citep{schmidt2019recurrent}.



\section{Conclusions}\label{chap:conclusion}

In this work we employed machine learning methods, and in particular neural networks, to predict baryonic properties of galaxies from dark matter data. Our networks learned from hydrodynamic simulations. 
We made predictions for stellar mass with a fairly good performance, but when predicting the SFR of galaxies the results were far from ideal. In both cases, predictions also presented undesirable behaviours at the lower and upper limits of the input values. However, the main results in this work are not linked with the quality of the predictions themselves, but on how the predictions relate with the chosen input variables.

Our results showed that stellar mass is a consistently predictable variable. Apart from the troublesome results at low and high values, we were able to reproduce to a fair extent the stellar mass function, the distribution of stellar masses with respect to halo mass, and how much they deviate from the mean of this relation. We outperformed the metrics achieved in other studies, both in stellar mass and SFR predictions. But, unlike stellar mass, SFR was not as well predicted. With the goal of reducing the scatter of this latter property, we used a time-averaged SFR, which is also a more meaningful property given that observational estimates of SFR are time averaged to different degrees. We highlight this time-averaging approach since it has not been adopted in previous works. This averaging improved our results in predicting SFR to some extent. We also achieved better results by using as input the output of a second neural network that reproduced stellar mass robustly, showing an interesting approach of linking networks to dig deeper in the information that can be deduced from our data. While there was a resemblance in the relation between SFR and stellar mass, the values predicted by the machine fall in a smaller range of values than the ones from TNG. Also, the deviation from the mean relation was, arguably, weakly recovered. All this points towards the need for more data, or different approaches in order to produce a consistent SFR prediction.

With respect to the relevance of input variables, we highlight in first place $z_{last}$ and $z_{1/2}$, two variables which have an important impact on the results of the predictions of both  baryonic properties. Specifically for SFR, $\dot{m}_{subhalo}$ also proved to be a feature of importance. This shows the relevance of considering events in the evolution of a galaxy and its environment when describing or predicting its baryonic properties. We also highlight the leave-one-covariate-out approach used to study the relevance of input variables, which is different to the methods used in similar studies.  Our implementation shows insights about how some relevant variables lose impact when added to the model with other highly correlated variables. 
Our results support the idea of not only considering instantaneous properties, which is also being considered in other related works. 

When analyzing the importance of input features, we also found that $v_{max}$ has a stronger impact than $m_{halo}$ in the prediction of $M^*$, consistent with previous analyses (eg. \citealt{Kulier}) but obtained independently by the machine which was able to infer that $v_{max}$ has more information pertinent to the stellar content of galaxies. This is interesting to highlight since, for example, when we study the correlation between input and output variables for $M^*$ (as seen in figure \ref{fig:corr_x}) we can see that the most correlated feature is $m_{halo}$, but this variable ranks fifth in importance. We can also see that variables with high importance, like $z_{half}$ in the case of $meanSFR$, is not highly correlated with the output variable. These and other insights from the comparison between correlation and impact in the model are relevant to address in future studies. 
The analysis itself of how each input variable affects the results can be of special interest and shed light on interesting relations. This could help, for example, to improve SAMs by hinting which properties should be given special attention. While it is a valid goal to use machine learning methods to populate dark matter simulations with less computational cost than required to run SAMs, these methods can also support and complement galaxy formation modelling.

Finally, taking all points in consideration, we conclude that machine learning models are not only strong candidates for potential simulation methods, but they also give us new tools and perspectives to understand the significance of different properties and their impact on target characteristics, enabling us to improve the already existing methods for recreating and studying the evolution of the universe and the Galaxies that live within it.

\section*{Acknowledgements}

We thank the Referee who significantly helped improve the clarity of this manuscript. CH acknowledges support from Fondecyt Regular 1191813, ANID, Chile.  NP was supported in part by a RAICES  and a PICT-2021-0700 grants from the Ministerio de Ciencia Tecnología e Innovación of Argentina.  

\section*{Data Availability}

No new data were generated in this work.



\bibliographystyle{mnras}
\bibliography{bibtex} 








\bsp	
\label{lastpage}
\end{document}